\documentclass[10pt,a4paper]{article}
\usepackage[utf8]{inputenc}
\usepackage[english]{babel}
\usepackage{amsmath}
\usepackage{amsfonts}
\usepackage{amssymb}
\usepackage{mathtools}
\usepackage{graphicx}
\usepackage{hyperref}
\usepackage{subfig}
\usepackage{overpic}
\usepackage{csquotes}
\usepackage[sorting=none]{biblatex}
\newcommand{\mean}[1]{\langle #1 \rangle}
\newcommand{\smat}[1]{\begin{smallmatrix} #1 \end{smallmatrix}}
\newcommand{\psmat}[1]{\left(\begin{smallmatrix} #1 \end{smallmatrix}\right)}
\newcommand{\bsmat}[1]{\left[\begin{smallmatrix} #1 \end{smallmatrix}\right]}
\newcommand{\pmat}[1]{\begin{pmatrix} #1 \end{pmatrix}}

\newcommand{\defeq}[0]{\vcentcolon=}
\newcommand{\eqdef}[0]{=\vcentcolon}
\newcommand{\pd}{\mathrm{d}}
\newcommand{\ticksize}{\footnotesize}

\title{Symmetry-Breaking and Hysteresis\\ in a Duplex Voter Model}
\author{Christian Kluge\footnote{Technical University of Munich, School of Computation Information and Technology, Department of Mathematics, Garching b.~M\"unchen, Germany}~ and Christian Kuehn\footnotemark[\value{footnote}]~\footnote{Munich Data Science Institute, Garching b.~M\"unchen, Germany}~\footnote{Complexity Science Hub Vienna, Vienna, Austria}}

\begin{document}

\maketitle

\begin{abstract}
    We introduce and analyze a voter-type model on a two-layer multiplex network, where the presence of a state on one layer acts as a catalyst or inhibitor to the propagation of that state on the other layer.
    Despite the model's simplicity, our mathematical analysis reveals a rich phase diagram that includes spontaneous symmetry-breaking and a cusp bifurcation, which arises when noise is introduced into the model. In particular, this bifurcation mechanism can be viewed as a prototypical unfolding of the change between explosive and non-explosive transitions observed in various other network models. We cross-validate our analytic results by numerical simulations.
\end{abstract}

\section{Introduction}

The voter model \cite{Clifford1973, Holley1975} has been a classic model of collective behavior since its introduction in the 1970s, enjoying particular popularity in opinion dynamics and evolutionary dynamics.
Over the decades, numerous variations of the model have been explored, such as the introduction of bias, noise, network rewiring, multiple states, or agents with special behavior.
See \cite{Dong2018, Redner2019} for an overview of these developments.

As the network science community realized the importance of multiple kinds of interactions in many real systems, there has been a veritable explosion of research on so-called \textit{multilayer} (or multiplex) models of collective dynamics \cite{Kivela2014, Boccaletti2014, Aleta2025}.
This has included multilayer versions of the voter model, most of which consider the spread of a single contagion across multiple kinds of edges \cite{Diakonova2014, Diakonova2016, Min2019}.
But there has also been interest in the concurrent spreading of multiple contagions.
Notably, Sanz et al.~\cite{Sanz2014} examined a coupling between two epidemic processes on a multiplex network, where the infection and recovery rates of vertices in one process depend on the state of those vertices in the other process.

We build on their line of inquiry by introducing and analyzing a novel variant of the voter model on a two-layer multiplex network, where the state on one layer modulates the propagation on the other.
In our model, vertices have a separate binary state ($A$ or $B$) on each layer.
The states on each layer evolve according to a biased edge voter model \cite{Lieberman2005} (also known as \textit{link dynamics} \cite{Antal2006, Sood2008}), where neighboring vertices \textquote{convince} each other through purely pairwise interactions.
The layers are coupled: when a vertex has state B on one layer, its transition rate to state B on the other layer is modified.
Positive coupling gives a catalytic behavior, negative coupling makes it inhibitory.
Additionally, the model incorporates noise in the form of spontaneous state changes.

We employ a mean-field approach to derive tractable equations governing the system's dynamics, and test the validity of our analytical predictions through numerical simulations.

Despite the edge voter model's reputation as being rather simple, we find that the coupling gives rise to a rich phase diagram, including phases of bistability and symmetry-breaking.
We also observe that small levels of noise unfold degenerate bifurcations into generic ones, revealing a cusp bifurcation that governs a transition between explosive and non-explosive (or sub- and super-critical, or first and second order) phase changes.
Through simulations on Erdős-Rényi, Barabási-Albert and lattice networks, we evaluate the robustness and limitations of our mean-field approach, showing that it accurately predicts the qualitative behavior on heterogeneous networks, but does not work equally well in the presence of layer overlap and short loops.

The remainder of this paper is structured as follows.
In section \ref{sec:modeldef}, we define the model, for which we then perform the mean-field analysis in section \ref{sec:analysis}.
The comparison to simulations is done in section \ref{sec:simulations}.

\section{Model Definition}\label{sec:modeldef}
We consider a two-layer multiplex network, given by a finite set $V$ of vertices and two sets $E_1,E_2$ of undirected edges.
We denote the number of vertices by $N$ and use $v \overset{_1}{\sim} w$ $\left(v \overset{_2}{\sim} w\right)$ to denote the existence of an edge in $E_1$ $\left(E_2\right)$ that connects $v,w\in V$.

In our model, each vertex has a state on each layer, and these states are either $A$ or $B$.
We denote the state of the whole system at time $t\in [0,\infty)$ by
\[\eta_t: {V\to \big\{\psmat{A\\A},\psmat{A\\B},\psmat{B\\A},\psmat{B\\B}\big\},} {\ v \mapsto \eta_t(v)}.\]
Here, the upper entry of $\eta_t(v)$ is the state of $v\in V$ on the first layer, and the lower entry is the state on the second layer.
We use $\eta_t^{(1)}(v)$ or $\eta_t^{(2)}(v)$ to separately refer to the state of $v$ on the first layer or on the second layer.

We now describe the separate dynamics on each layer.
The states on both layers evolve according to the same biased edge voter model, where state-$A$-vertices \textquote{convince} their neighbors with rate $\alpha$, while state-$B$-vertices convince their neighbors with a different rate $\beta$.
Additionally, vertices flip their state with rate $\varepsilon \ge 0$ independently of their neighbors, to introduce noise into the system.

These dynamics on the two layers are coupled by a similar mechanism as those explored by Sanz et al.~\cite{Sanz2014} in the context of SIS and SIR epidemic models on multiplex networks.
If a vertex has state $B$ on one layer and state $A$ on the other, then the parameter $\beta$ for that vertex is replaced by $\beta(1+\delta)$, where $\delta\in[-1,\infty)$ is a parameter determining the strength and quality of the coupling.
For $\delta>0$, the presence of state $B$ on one layer acts as a catalyst for the spread of $B$ on the other layer, while for $\delta<0$ it acts as an inhibitor.

In summary, the possible transitions and their rates are as follows on the top layer, and analogous on the bottom layer:
\begin{align*}
    \smat{B&-&A\\&&&} & \overset{\alpha}{\xrightarrow[\hspace{1cm}]{}} \smat{A&-&A\\&&&}\\
    \smat{A&-&B\\A&&} & \overset{\beta}{\xrightarrow[\hspace{1cm}]{}} \smat{B&-&B\\A&&}\\
    \smat{A&-&B\\B&&} & \overset{(1+\delta)\beta}{\xrightarrow[\hspace{1cm}]{}} \smat{B&-&B\\B&&}\\
    \smat{A\\&}       & \overset{\varepsilon}{\xrightleftharpoons[\hspace{1cm}]{}}  \smat{B\\&}
\end{align*}

To be precise, this is a continuous-time Markov chain, where $\eta_t^{(L)}(v)=B$ ($L\in\{1,2\}$) transitions to $A$ with rate
\[
\varepsilon + \left|\left\{w\in V \mid w \overset{_L}{\sim} v ,\ \eta_t^{(L)}(w)=A \right\}\right| \cdot\alpha,
\]
and $\eta_t^{(L)}(v)=A$ transitions to $B$ with rate
\[
\varepsilon + \left|\left\{w\in V \mid w \overset{_L}{\sim} v ,\ \eta_t^{(L)}(w)=B \right\}\right| \cdot\beta\cdot
\begin{cases}
    (1+\delta),&\text{ if } \eta_t^{(3-L)}(v)=B\\
    1,&\text{ else}
\end{cases}
.
\]

\section{Mean-Field Analysis}\label{sec:analysis}
\subsection{Derivation of the Mean-Field Equation}
We wish to obtain a low-dimensional approximation of this model that nonetheless captures the most important dynamics.
To this end, we aim at a particular set of observables: The number of vertices in each possible state.
Let $\bsmat{X\\Y}$ ($X,Y\in\{A,B\}$) denote the number of vertices in state $\psmat{X\\Y}$.
This is a random variable whose value depends on time.
At all times these variables satisfy the equation
\[
\bsmat{A\\A} + \bsmat{A\\B} + \bsmat{B\\A} + \bsmat{B\\B} = N.
\]
We use this equation to express $\bsmat{A\\A}$ as $N-\bsmat{A\\B}-\bsmat{B\\A}-\bsmat{B\\B}$, thus eliminating the need to explicitly keep track of it.
Next, we replace the set of observables $\bsmat{A\\B},\bsmat{B\\A},\bsmat{B\\B}$ by the equivalent
$\bsmat{\phantom{A}\\B},\bsmat{B\\\phantom{A}},\bsmat{B\\B}$,
where ${\bsmat{\phantom{A}\\B} \defeq \bsmat{A\\B}+\bsmat{B\\B}}$ and $\bsmat{B\\\phantom{A}} \defeq \bsmat{B\\A}+\bsmat{B\\B}$.
These observables have a nice interpretation:
$\bsmat{B\\\phantom{A}}$ is the number of vertices where $\eta_t^{(1)}=B$,
$\bsmat{\phantom{A}\\B}$ is the number of vertices where $\eta_t^{(2)}=B$,
and $\bsmat{B\\B}$ is the size of the overlap between these two sets of vertices.
Our goal is to understand the evolution of the expected values of $\bsmat{B\\\phantom{A}}$ and $\bsmat{\phantom{A}\\B}$, since these tell us whether state $A$ or $B$ is in the majority in each layer.

We follow a moment closure approach to obtain a low-dimensional differential equation describing the evolution of these quantities.
As moments we choose expected counts of certain subgraph states, often referred to as \textit{network motifs}.
We use the following notation for these moments, where $X,Y,Z \in \{A,B\}$:
\begin{itemize}
    \item $\mean{\smat{X\\Y}}$: the expected number of vertices with state $X$ on layer 1 and state $Y$ on layer 2.
    \item $\mean{\smat{X\\\phantom{B}}}$: the expected number of vertices with state $X$ on layer 1, regardless of the state on layer 2.
    \item $\mean{\smat{\phantom{B}\\X}}$: the expected number of vertices with state $X$ on layer 2, regardless of the state on layer 1.
    \item $\mean{\smat{X&Y\\\phantom{B}&}}$: the expected number of $X-Y$ edges on layer 1.
    \item $\mean{\smat{\phantom{B}&\\X&Y}}$: the expected number of $X-Y$ edges on layer 2.
    \item $\mean{\smat{X&Z\\Y&}}$: the expected number of layer-1-edges from an $\psmat{X\\Y}$-vertex to a vertex with state $Z$ on layer 1, regardless of the layer-2-state of that vertex.
    \item $\mean{\smat{X&\\Y&Z}}$: the expected number of layer-2-edges from an $\psmat{X\\Y}$-vertex to a vertex with state $Z$ on layer 2, regardless of the layer-1-state of that vertex.
\end{itemize}
All these quantities depend on time, but we do not make this dependence explicit in the notation.

We can heuristically obtain differential equations for $\mean{\smat{B\\\phantom{B}}}$ and $\mean{\smat{\phantom{B}\\B}}$ as follows:
A $\psmat{B\\\phantom{B}}$-vertex is created either from an $\psmat{A\\\phantom{B}}$-vertex through noise with rate $\varepsilon$,
or from an $A-B$ edge with rate $\beta$ or $\beta(1+\delta)$, depending on the layer-2 state of the vertex in question.
On the other hand, $\psmat{B\\\phantom{B}}$-vertices are lost either through noise with rate $\varepsilon$, or by being in a $B-A$ edge with rate $\alpha$.

Taken together, this leads us to the equation
\begin{align*}
\frac{\pd}{\pd t} \mean{\smat{B\\\phantom{B}}} =
    &+\varepsilon \mean{\smat{A\\\phantom{A}}}
    + \beta \mean{\smat{A&B\\A&}} + \beta(1+\delta) \mean{\smat{A&B\\B&}}\\
    &-\varepsilon \mean{\smat{B\\\phantom{B}}}
    -\alpha \mean{\smat{A&B\\&}}.
\end{align*}
By the same reasoning we obtain an analogous equation for $\mean{\smat{\phantom{B}\\B}}$.
We can then simplify these equations using the fact that $\mean{\smat{A&B\\A&}}+\mean{\smat{A&B\\B&}} = \mean{\smat{A&B\\&}}$
and arrive at
\begin{align*}
    \frac{\pd}{\pd t} \mean{\smat{B\\\phantom{B}}} &= \varepsilon (\mean{\smat{A\\\phantom{A}}}-\mean{\smat{B\\\phantom{B}}})
                                                + (\beta-\alpha)\mean{\smat{A&B\\&}} + \beta\delta \mean{\smat{A&B\\B&}} \\
    \frac{\pd}{\pd t} \mean{\smat{\phantom{B}\\B}} &= \varepsilon (\mean{\smat{\phantom{A}\\A}}-\mean{\smat{\phantom{B}\\B}})
                                                + (\beta-\alpha)\mean{\smat{&\\A&B}} + \beta\delta \mean{\smat{B&\\A&B}}.
\end{align*}
This system of equations is not \textquote{closed} in the sense that it contains moments whose evolution it does not specify, e.g.~$\mean{\smat{A&B\\B&}}$.
One could tackle this by also including equations for these moments, but those equations will necessarily contain new higher-order moments and the system will still not be closed.
Instead, we apply a \textit{moment closure}, which is the usual solution to this problem \cite{KuehnMC,Kiss2017}.
That means we approximate the \textquote{problematic} moments using the lower-order moments that are described by the equations.

Let $N$ be the number of vertices in the network, and $k_1,k_2$ the average degree on layer 1 and 2, respectively.
One of the closures we choose is the usual pair closure
\begin{equation}\label{eq:closure-pair}
    \mean{\smat{A&B\\&}} \approx \frac{k_1}{N} \mean{\smat{A\\\phantom{A}}} \mean{\smat{B\\\phantom{B}}}.
\end{equation}
It can easily be derived by assuming that each $A$-vertex is connected to $k_1$ other vertices, which independently have probability $\frac{1}{N}\mean{\smat{B\\\phantom{B}}}$ to be in state $B$.
The other closure we use is
\begin{equation}\label{eq:closure-other}
    \mean{\smat{A&B\\B&}} \approx \frac{k_1}{N} \mean{\smat{A\\\phantom{A}}} \mean{\smat{B\\\phantom{B}}} \frac{\mean{\smat{\phantom{B}\\B}}}{N},
\end{equation}
which is an extension of the previous one and assumes that the $A$-vertex in question is an $\psmat{A\\B}$-vertex with probability $\frac{1}{N}\mean{\smat{\phantom{B}\\B}}$.

We apply these closures and the relation $\mean{\smat{A\\\phantom{A}}}+\mean{\smat{B\\\phantom{B}}}=N$
to obtain the following closed system of equations, where $b_1\defeq \frac{1}{N}\mean{\smat{B\\\phantom{B}}},b_2\defeq \frac{1}{N}\mean{\smat{\phantom{B}\\B}}$:
\begin{align*}
    \dot b_1 &= N\varepsilon (1-2 b_1) + k_1 N b_1 (1-b_1)(\beta-\alpha+\beta\delta b_2) \\
    \dot b_2 &= N\varepsilon (1-2 b_2) + k_2 N b_2 (1-b_2)(\beta-\alpha+\beta\delta b_1)
\end{align*}
Note that the domain of this equation is the unit square $[0,1]^2$, since \mbox{$b_1,b_2\in[0,1]$} describe the prevalence of state $B$ relative to the total number of vertices.
We will refer to $[0,1]^2$ as the \textquote{physically meaningful region} to distinguish it from the rest of $\mathbb{R}^2$.

Before we tackle the analysis, we want to simplify this equation further.
First, we note that we are only interested in the long-term behavior of the model, and are thus free to re-scale time.
Second, we impose the assumption that $k_1=k_2\eqdef k$.
This restricts the generality of the model, since it forces us to only consider networks where the average degrees on both layers are equal.
However, the model is already quite interesting on such networks and the analysis is made much easier by this restriction.
We now apply a scaling given by
\begin{align}\label{eq:rescaling}
    \tilde t \defeq & k N \beta \cdot t, &
    \tilde \varepsilon \defeq & \frac{\varepsilon}{k\beta}, &
    \tilde \alpha \defeq & \frac{\alpha}{\beta}.
\end{align}
We then drop the tildes for the sake of readability, but we keep in mind that we will be working with the transformed variables for the remainder of the analysis.
We thus arrive at our final mean-field equation, which we will analyze from now on:
\begin{equation}\label{eq:2dmf}
\begin{split}
    \dot b_1 &= \varepsilon (1-2 b_1) + b_1 (1-b_1)(1-\alpha+\delta b_2) \\
    \dot b_2 &= \varepsilon (1-2 b_2) + b_2 (1-b_2)(1-\alpha+\delta b_1)
\end{split}\end{equation}
As an aside, let us remark that for $\varepsilon=0$ and $\delta=0$, these are simply two separate replicator equations where species $A$ has fitness $\alpha$ and species $B$ has fitness $1$.
For $\delta > 0$, the fitnesses of the two $B$ species are coupled in a manner resembling symbiosis.

\subsection{Analysis without Noise}\label{sec:ana-no-noise}
We first focus our analysis of the mean-field equation on the case without noise, i.e.~$\varepsilon=0$.
For generic parameters (i.e.~$1\ne \alpha\ne 1+\delta$), the equation is simple enough that we can easily read off the equilibria
\[
\pmat{0\\0},\pmat{1\\0},\pmat{0\\1},\pmat{1\\1},\pmat{b^\star\\b^\star},
\]
where $b^\star \defeq \frac{\alpha-1}{\delta}$. Linearizing at these equilibria leads to an eigenvalue problem for eigenvalues $\lambda_{1,2}$, which determine the local stability.
To work with these eigenvalues, it will be helpful to define
\[
    \kappa_1\defeq 1-\alpha \qquad\text{and}\qquad \kappa_2\defeq -(1-\alpha+\delta).
\]
Solving the eigenvalue problem for $\psmat{0\\0}$ and $\psmat{1\\1}$ then yields that both of these equilibria are vertices that can be stable or unstable, depending on the signs of $\kappa_1$ and $\kappa_2$.
For $\psmat{0\\0}$ we have $\lambda_1 = \lambda_2 = \kappa_1$,
and for $\psmat{1\\1}$ we have $\lambda_1 = \lambda_2 = \kappa_2$.
$\psmat{b^\star\\b^\star}$ is a saddle point with eigenvectors $\psmat{1\\1},\psmat{-1\\1}$
and eigenvalues $\lambda_1 = - \kappa_1 (1-b^\star)$ and $\lambda_2 = + \kappa_1 (1-b^\star)$.
We note that since $b^\star=\frac{\kappa_1}{\kappa_1+\kappa_2}$, this equilibrium exists in the physically meaningful region $[0,1]^2$ if and only if $\kappa_1$ and $\kappa_2$ have the same sign.
$\psmat{1\\0}$ and $\psmat{0\\1}$ can be either nodes or saddle points, depending on the parameters.
$\psmat{1\\0}$ has the eigenvectors $\psmat{1\\0},\psmat{0\\1}$ with eigenvalues $\lambda_1=-\kappa_1,\, \lambda_2=-\kappa_2$.
$\psmat{0\\1}$ has the same eigenvectors $\psmat{1\\0},\psmat{0\\1}$, but with eigenvalues $\lambda_1=-\kappa_2,\, \lambda_2=-\kappa_1$.

These pieces of local information already give us a good picture of the global behavior of the system, since we can rule out the existence of periodic orbits by the following argument:
In a planar ODE, any periodic orbit needs to encircle at least one equilibrium (see Prop.~1.8.4 in \cite{Guckenheimer1983}), which is impossible here.
Four of the equilibria are located on the boundary of the domain $[0,1]^2$ and it is easy to see from equation \eqref{eq:2dmf} that each of the four sides of the boundary is an invariant set (for $\varepsilon=0$), thus making it impossible for any orbit to cross them.
The remaining equilibrium is located on the diagonal $\{\psmat{b\\b} \mid b\in [0,1]\}$, which is also an invariant set as can easily be seen from the equation.

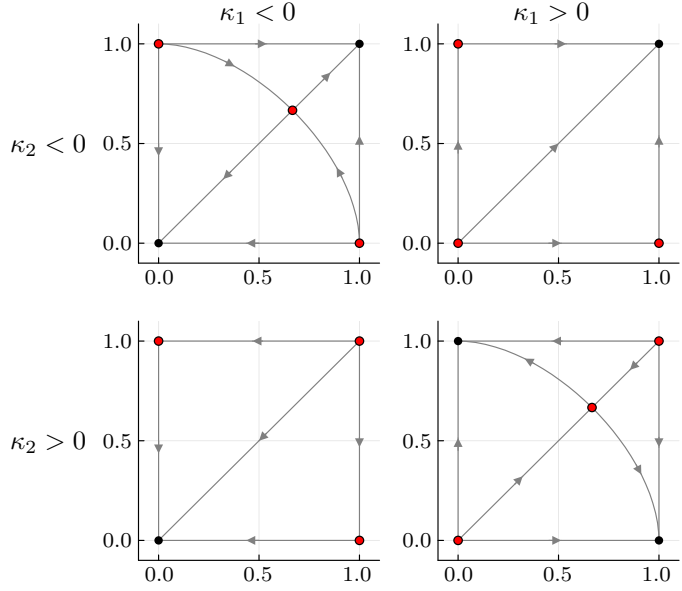
\begin{figure}
    \centering
    \begin{overpic}[width=0.65\linewidth]{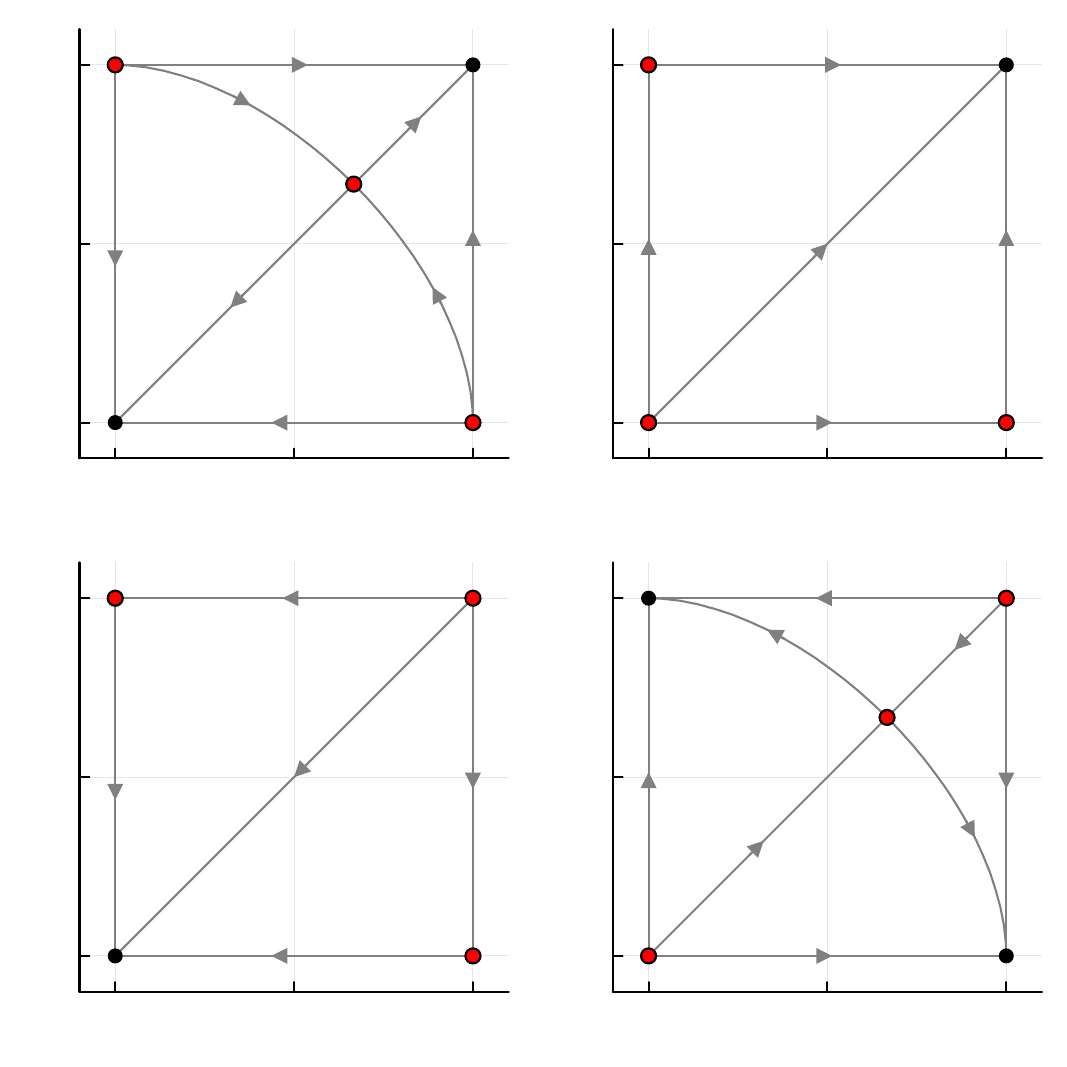}
        \put(21,98){$\kappa_1<0$}
        \put(70,98){$\kappa_1>0$}
        \put(-14,25.75){$\kappa_2>0$}
        \put(-14,75.75){$\kappa_2<0$}
        
        \put(1.5,9.25){\ticksize $0.0$}
        \put(1.5,26){\ticksize $0.5$}
        \put(1.5,42.75){\ticksize $1.0$}
        
        \put(1.5,59.25){\ticksize $0.0$}
        \put(1.5,76){\ticksize $0.5$}
        \put(1.5,92.75){\ticksize $1.0$}

        \put(51.5,9.25){\ticksize $0.0$}
        \put(51.5,26){\ticksize $0.5$}
        \put(51.5,42.75){\ticksize $1.0$}

        \put(51.5,59.25){\ticksize $0.0$}
        \put(51.5,76){\ticksize $0.5$}
        \put(51.5,92.75){\ticksize $1.0$}

        \put(8.5,3.5){\ticksize $0.0$}
        \put(25,3.5){\ticksize $0.5$}
        \put(41.5,3.5){\ticksize $1.0$}

        \put(58.5,3.5){\ticksize $0.0$}
        \put(75,3.5){\ticksize $0.5$}
        \put(91.5,3.5){\ticksize $1.0$}

        \put(8.5,53.5){\ticksize $0.0$}
        \put(25,53.5){\ticksize $0.5$}
        \put(41.5,53.5){\ticksize $1.0$}

        \put(58.5,53.5){\ticksize $0.0$}
        \put(75,53.5){\ticksize $0.5$}
        \put(91.5,53.5){\ticksize $1.0$}
    \end{overpic}
    \caption{Phase portraits representative of the four possible phases. In the left two portraits we have $\kappa_1=-\frac 1 2$, on the right we have $\kappa_1=+\frac 1 2$. The top portraits have $\kappa_2=-\frac 1 4$, the bottom portraits have $\kappa_2=+\frac 1 4$.}
    \label{fig:phase_portraits}
\end{figure}

Depending on the signs of $\kappa_1$ and $\kappa_2$, there can be four qualitatively different phase portraits, which are illustrated in Fig.~\ref{fig:phase_portraits}.
We consider these to be the different phases of the system:
\begin{itemize}
    \item The low-B phase, characterized by $\kappa_1<0,\kappa_2>0$, where $\psmat{0\\0}$ is the unique globally stable equilibrium.
        In this phase, all vertices will end up in state A on both layers.
    \item The high-B phase, characterized by $\kappa_1>0,\kappa_2<0$, where $\psmat{1\\1}$ is the unique globally stable equilibrium.
        In this phase, all vertices will end up in state B on both layers.
    \item The bistable phase, characterized by $\kappa_1<0,\kappa_2<0$, where both $\psmat{0\\0}$ and $\psmat{1\\1}$ are locally stable, and their basins of attraction are separated by the saddle point $\psmat{b^\star\\b^\star}$ and by the heteroclinic orbits connecting it to $\psmat{1\\0}$ and $\psmat{0\\1}$.
        In this phase, state B may go extinct on both layers or dominate on both layers, depending on the initial condition.
    \item The symmetry-breaking phase, characterized by $\kappa_1>0,\kappa_2>0$, where $\psmat{1\\0}$ and $\psmat{0\\1}$ are the stable equilibria, and their basins of attraction are separated by the diagonal $\{\psmat{b\\b}\mid b\in[0,1]\}$.
        In this phase, state B will go extinct on one layer and dominate on the other, despite the model being defined symmetrically with respect to the layers.
\end{itemize}

These phases are separated by bifurcations, which occur when one of $\kappa_1,\kappa_2$ passes through $0$.
As one might expect from the eigenvalues of the equilibria, these bifurcations are somewhat degenerate.
As $\kappa_1$ passes through $0$ (corresponding to a passage from the left to the right in Fig.~\ref{fig:phase_portraits}), the equilibria $\psmat{0\\0}$ and $\psmat{b^\star\\b^\star}$ pass through each other and exchange stability.
Morally speaking, this is of course a transcritical bifurcation, and one would generically expect that there exists a one-dimensional center eigenspace at the bifurcation point.
However, that is not the case here, as the flow on the adjacent segments of the boundary of $[0,1]^2$ reverses its direction simultaneously, and at $\kappa_1=0$ all points in  $\{\psmat{b_1\\b_2} \mid b_1=0\ \vee\ b_2=0\}$ are fixed points, resulting in a two-dimensional center eigenspace.
In addition, this flow reversal on the boundary also changes the stability of the equilibria $\psmat{1\\0}$ and $\psmat{0\\1}$.
Similarly, as $\kappa_2$ passes through $0$ (corresponding to passage between the rows of Fig.~\ref{fig:phase_portraits}), the equilibrium $\psmat{b^\star\\b^\star}$ collides with $\psmat{1\\1}$ 
and we again have a transcritical bifurcation with the same degeneracy.
This time the lines on which the flow reverses are $\{\psmat{b_1\\b_2} \mid b_1=1\ \vee\ b_2=1\}$,
and the stabilities of $\psmat{b^\star\\b^\star},\psmat{1\\1},\psmat{1\\0}$ and $\psmat{0\\1}$ again change all at once.
When both $\kappa_1=0$ and $\kappa_2=0$, the right-hand side of the mean-field \eqref{eq:2dmf} vanishes entirely.

\subsection{Analysis with Small Noise}
For $\varepsilon>0$, we are no longer able to easily obtain closed formulas for the equilibria, making analysis more difficult.
However, we can use a perturbation approach to compute the derivatives of the equilibria with respect to $\varepsilon$ at $\varepsilon=0$.
To do so, we make the ansatz $b = b^{(0)} + b^{(1)} \varepsilon$, where $b^{(0)}$ is one of the known equilibria from the unperturbed case, $\varepsilon$ is nonzero but we formally assume $\varepsilon^2=0$, and we solve $\dot b = 0$ for $b^{(1)}$.
So for $b^{(0)} = \psmat{0\\0}$ the ansatz to find $b^{(1)}_1$ is
\begin{align*}
   0 = \dot b_1 &= \varepsilon (1-2(0+b^{(1)}_1\varepsilon)) + (0+b^{(1)}_1\varepsilon) (1-(0+b^{(1)}_1\varepsilon))(1-\alpha+\delta (0+b^{(1)}_2\varepsilon))\\
   &= \varepsilon + \varepsilon \cdot b^{(1)}_1 (1-\alpha),
\end{align*}
which implies $b^{(1)}_1 = -\frac{1}{1-\alpha}=-\frac{1}{\kappa_1}$. An analogous calculation yields $b^{(1)}_2=-\frac{1}{\kappa_1}$.

We find:
\begin{itemize}
    \item $b^{(0)} = \psmat{0\\0} \implies b^{(1)} = -\frac{1}{\kappa_1}\psmat{1\\1}$:
        This equilibrium moves upwards along the diagonal if it is stable, and leaves the physically meaningful region $[0,1]^2$ if unstable.
    \item $b^{(0)} = \psmat{1\\1} \implies b^{(1)} = +\frac{1}{\kappa_2}\psmat{1\\1}$:
        This equilibrium moves down along the diagonal if it is stable, and leaves the physically meaningful region $[0,1]^2$ if unstable.
    \item $b^{(0)} = \psmat{1\\0} \implies b^{(1)} = \psmat{-\frac{1}{\kappa_1}\\+\frac{1}{\kappa_2}}$:
        This equilibrium moves into the interior of $[0,1]^2$ in the symmetry-breaking phase. In the other phases, it will leave $[0,1]^2$.
    \item $b^{(0)} = \psmat{0\\1} \implies b^{(1)} = \psmat{+\frac{1}{\kappa_2}\\-\frac{1}{\kappa_1}}$:
        This equilibrium moves into the interior of $[0,1]^2$ in the symmetry-breaking phase. In the other phases, it will leave $[0,1]^2$.
    \item $b^{(0)} = \psmat{b^\star\\b^\star} \implies b^{(1)} = \frac{2b^\star -1}{\delta b^\star (1-b^\star)}\psmat{1\\1}$:
        This equilibrium moves on the diagonal towards $\psmat{\frac{1}{2}\\ \frac{1}{2}}$ if $\delta<0$, and away from it if $\delta>0$.
\end{itemize}

\begin{figure}
    \centering
    \begin{overpic}[width=0.65\linewidth]{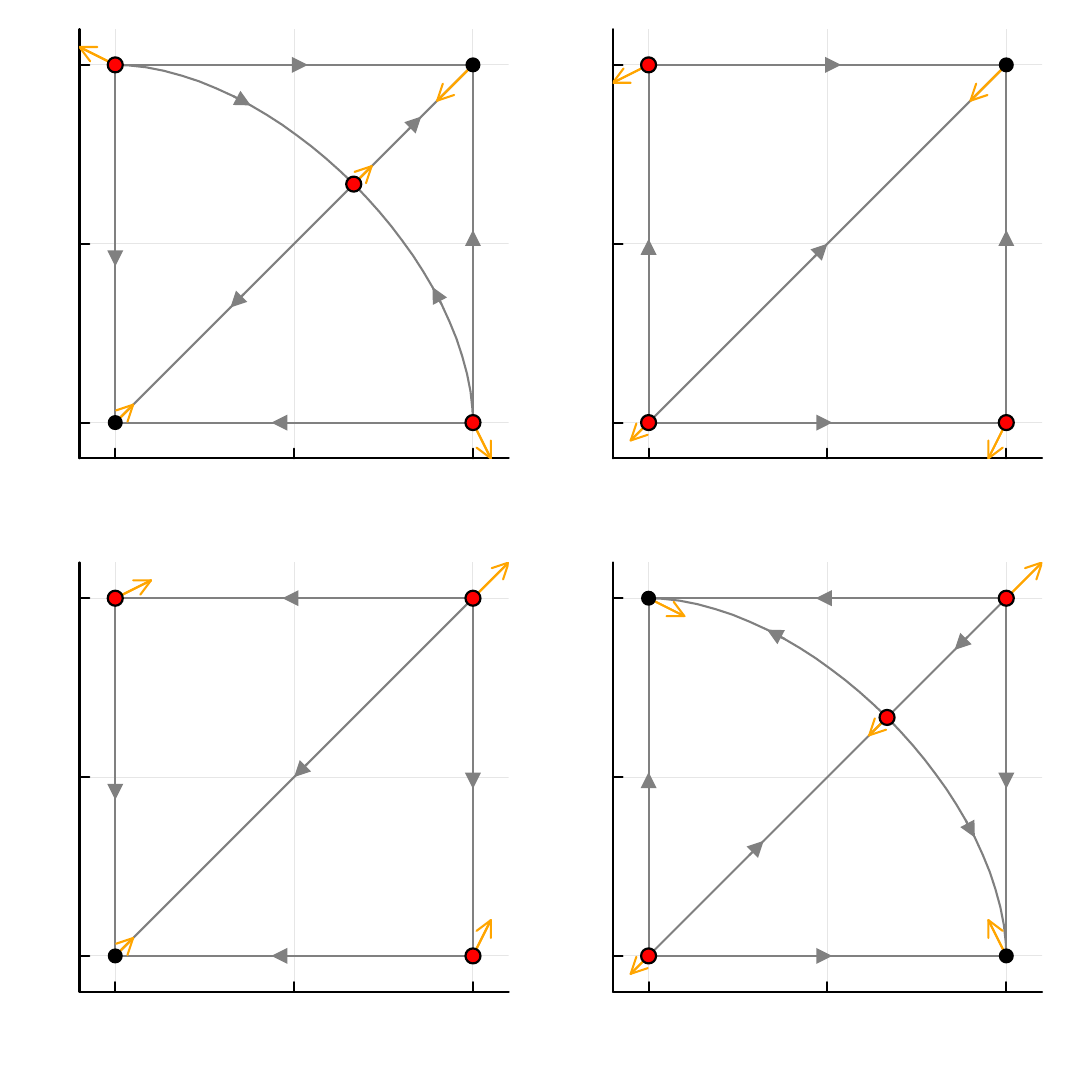}
        \put(21,98){$\kappa_1<0$}
        \put(70,98){$\kappa_1>0$}
        \put(-14,25.75){$\kappa_2>0$}
        \put(-14,75.75){$\kappa_2<0$}
        
        \put(1.5,9.25){\ticksize $0.0$}
        \put(1.5,26){\ticksize $0.5$}
        \put(1.5,42.75){\ticksize $1.0$}
        
        \put(1.5,59.25){\ticksize $0.0$}
        \put(1.5,76){\ticksize $0.5$}
        \put(1.5,92.75){\ticksize $1.0$}

        \put(51.5,9.25){\ticksize $0.0$}
        \put(51.5,26){\ticksize $0.5$}
        \put(51.5,42.75){\ticksize $1.0$}

        \put(51.5,59.25){\ticksize $0.0$}
        \put(51.5,76){\ticksize $0.5$}
        \put(51.5,92.75){\ticksize $1.0$}

        \put(8.5,3.5){\ticksize $0.0$}
        \put(25,3.5){\ticksize $0.5$}
        \put(41.5,3.5){\ticksize $1.0$}

        \put(58.5,3.5){\ticksize $0.0$}
        \put(75,3.5){\ticksize $0.5$}
        \put(91.5,3.5){\ticksize $1.0$}

        \put(8.5,53.5){\ticksize $0.0$}
        \put(25,53.5){\ticksize $0.5$}
        \put(41.5,53.5){\ticksize $1.0$}

        \put(58.5,53.5){\ticksize $0.0$}
        \put(75,53.5){\ticksize $0.5$}
        \put(91.5,53.5){\ticksize $1.0$}
    \end{overpic}
    \caption{The same as Fig.~\ref{fig:phase_portraits}, but with the addition of orange arrows indicating the directions in which the equilibria move for small $\varepsilon>0$, based on our perturbation calculation.}
    \label{fig:perturbation_portraits}
\end{figure}
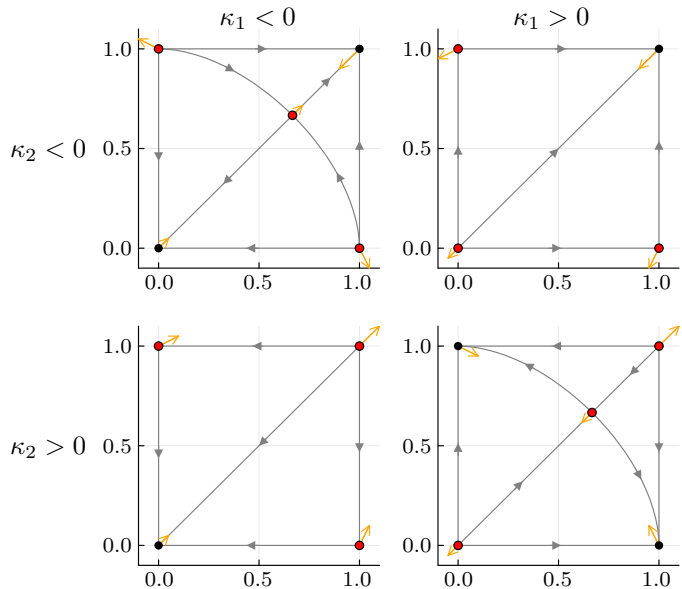

These findings are illustrated in Fig.~\ref{fig:perturbation_portraits}. Even a small value of $\varepsilon>0$ qualitatively changes the phase portraits,
as the unstable equilibria on the boundary leave the region $[0,1]^2$.
Beyond that, $\varepsilon$ also affects the bifurcations present in the system.
This can be seen easily when we consider the previously degenerate case $\kappa_1=\kappa_2=0$, where for $\varepsilon=0$ the right-hand-side of the mean-field \eqref{eq:2dmf} vanished: For any positive $\varepsilon$, there is now a unique equilibrium at $\psmat{\frac{1}{2}\\ \frac{1}{2}}$.
Unfortunately, this perturbation calculation does not give us information about how exactly the bifurcations are affected.
Additionally, it cannot tell us anything about the case of $\varepsilon \gg 0$, where we expect new bifurcations to arise.
In the next section we will use a different approach to gain more insight into both of these issues.

\subsection{Analysis Restricted to the Diagonal}\label{sec:ana-diagonal}
As we can see from Fig.~\ref{fig:phase_portraits}, the diagonal $D\defeq \{\psmat{b\\b} \mid b\in [0,1]\}$ occupies a special place in the phase portrait.
It separates the phase space $[0,1]^2$ into two halves, which have identical dynamics up to reflection across the diagonal.
Furthermore, the dynamics inside these halves must be consistent with the dynamics on their boundaries, which consist of $D$ and the boundary of $[0,1]^2$.
From equation \eqref{eq:2dmf} we can read off that, for $\varepsilon>0$, the boundary of $[0,1]^2$ is not an invariant set
and the flow across it points inward everywhere.
On the other hand, the diagonal $D$ \emph{is} an invariant set, and the dynamics on $D$ is far less trivial.
It will therefore pay off to understand the dynamics on $D$ in more detail.
These dynamics are given by the one-dimensional ODE
\begin{equation}\label{eq:1dmf}
    \dot b = \varepsilon (1-2b) + b(1-b)(1-\alpha+\delta b).
\end{equation}

Without noise, i.e.~for $\varepsilon=0$, we can simply read off the equilibria, which are $0,\ 1,$ and $b^\star \defeq -\frac{\kappa_1}{\delta}$, where we still use the notation $\kappa_1\defeq 1-\alpha$.
As before in system \eqref{eq:2dmf}, the third equilibrium crosses the first at $\kappa_1=0$ and the second at $\kappa_1=-\delta$.
However, unlike before, these are now standard transcritical bifurcations as we can see by the following calculation.
Let $f(b,\kappa_1) \defeq b(1-b)(\kappa_1+\delta b)$ denote the RHS of system \eqref{eq:1dmf}.
We then compute a Taylor expansion of $f$ in $b$ and $\kappa_1$ to second order, obtaining
\[
    f(b,\kappa_1) = \kappa_1 b + \delta b^2 + h.o.t.
\]
around $b=0, \kappa_1=0$ and
\[
    f(b,\kappa_1) = -(\kappa_1+\delta)(b-1) -\delta(b-1)^2 + h.o.t.
\]
around $b=1, \kappa_1=-\delta$.

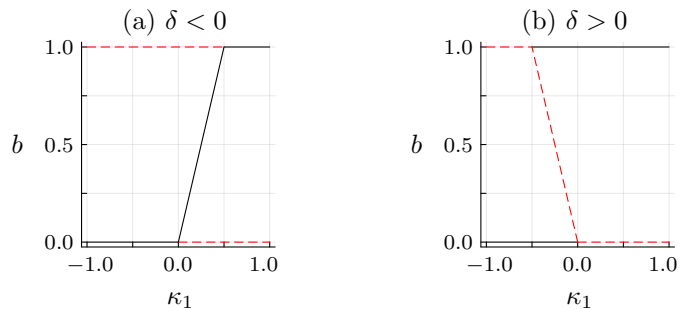
\begin{figure}
\centering
\subfloat{\begin{overpic}[width=0.3\linewidth]{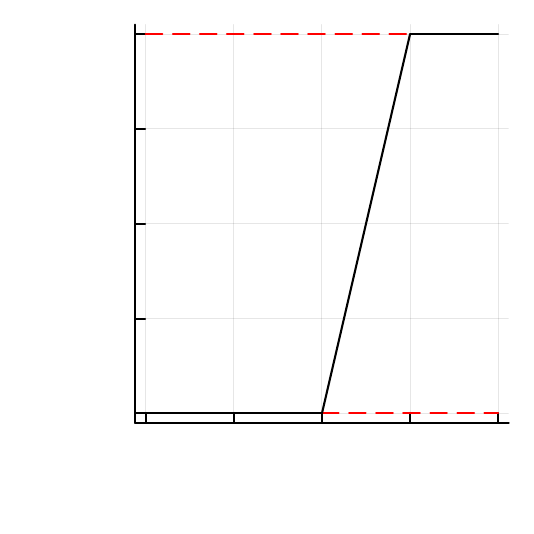}
    \put(40,100){(a) $\delta<0$}

    \put(0,55){$b$}
    \put(12,20){\ticksize $0.0$}
    \put(12,55.5){\ticksize $0.5$}
    \put(12,91){\ticksize $1.0$}

    \put(53,0){ $\kappa_1$}
    \put(20,12){\ticksize $-1.0$}
    \put(55,12){\ticksize $0.0$}
    \put(86,12){\ticksize $1.0$}
    \end{overpic}}
\hfil
\subfloat{\begin{overpic}[width=0.3\linewidth]{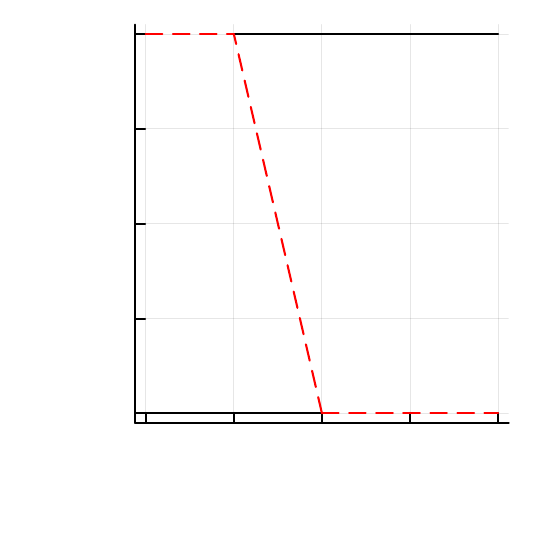}
    \put(40,100){(b) $\delta>0$}

    \put(0,55){$b$}
    \put(12,20){\ticksize $0.0$}
    \put(12,55.5){\ticksize $0.5$}
    \put(12,91){\ticksize $1.0$}

    \put(53,0){ $\kappa_1$}
    \put(20,12){\ticksize $-1.0$}
    \put(55,12){\ticksize $0.0$}
    \put(86,12){\ticksize $1.0$}
    \end{overpic}}
\caption{
Bifurcation diagrams showing the equilibria of system \eqref{eq:1dmf} with $\varepsilon=0$ as $\kappa_1$ is varied.
The solid lines are stable equilibria, the dashed lines are unstable equilibria.
The values of $\delta$ are $-\frac{1}{2}$ (a) and $+\frac{1}{2}$ (b).
We see the two transcritical bifurcations before and after the nonexplosive-to-explosive transition.
}
\label{fig:1d-bif-nonoise}
\end{figure}

Focusing on the bifurcation at $\kappa_1=0,b=0$, we thus have 
\[ \dot b \approx \kappa_1 b + \delta b^2, \]
which is the normal form of a transcritical bifurcation in $\kappa_1$.
The bifurcation is first-order for $\delta>0$ and second-order for $\delta<0$.
We therefore have a switch from a nonexplosive to an explosive phase transition,
which is what one expects when varying a transcritical bifurcation along an additional parameter without breaking it \cite{Kuehn2021}.
The other bifurcation for $\kappa_1=-\delta$ at $b=1$ behaves \textquote{in reverse} and undergoes the same change of criticality at $\delta=0$.
The bifurcations and their change in criticality are depicted in Fig.~\ref{fig:1d-bif-nonoise}.

Note however, that the stable intermediate branch of equilibria which exists for $\delta<0,\ \kappa_1\in(0,-\delta)$ is not observable in the global dynamics,
because it is located in the symmetry-breaking phase where the diagonal $D$ is unstable.
We still find this nonexplosive-to-explosive transition relevant, since for $\varepsilon>0$ it will turn into a cusp bifurcation, which is visible in the global dynamics.

\begin{figure}
\centering
\subfloat{\begin{overpic}[width=0.3\linewidth]{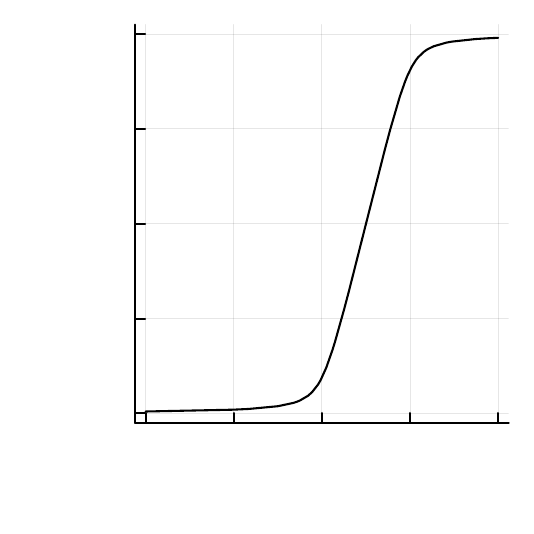}
    \put(40,100){(a) $\delta<0$}

    \put(0,55){$b$}
    \put(12,20){\ticksize $0.0$}
    \put(12,55.5){\ticksize $0.5$}
    \put(12,91){\ticksize $1.0$}

    \put(53,0){ $\kappa_1$}
    \put(20,12){\ticksize $-1.0$}
    \put(55,12){\ticksize $0.0$}
    \put(86,12){\ticksize $1.0$}
    \end{overpic}}
\hfil
\subfloat{\begin{overpic}[width=0.3\linewidth]{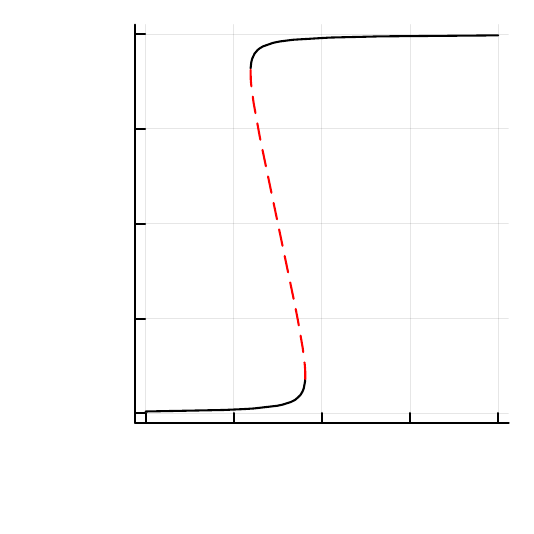}
    \put(40,100){(b) $\delta>0$}

    \put(0,55){$b$}
    \put(12,20){\ticksize $0.0$}
    \put(12,55.5){\ticksize $0.5$}
    \put(12,91){\ticksize $1.0$}

    \put(53,0){ $\kappa_1$}
    \put(20,12){\ticksize $-1.0$}
    \put(55,12){\ticksize $0.0$}
    \put(86,12){\ticksize $1.0$}
    \end{overpic}}
\caption{
Bifurcation diagrams showing the equilibria of system \eqref{eq:1dmf} with $\varepsilon=0.005$ as $\kappa_1$ is varied.
The solid lines are stable equilibria, the dashed lines are unstable equilibria.
The values of $\delta$ are $-\frac{1}{2}$ (a) and $+\frac{1}{2}$ (b).
}
\label{fig:1d-bif-noise}
\end{figure}
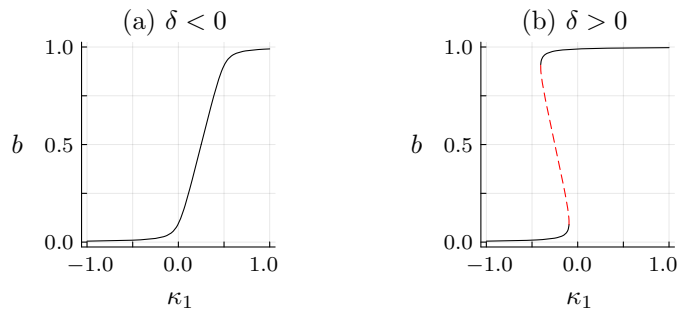

For small $\varepsilon>0$, each of the transcritical bifurcations unfolds into two separate branches of equilibria, of which only one lies inside the domain $[0,1]$. Additionally, in the explosive case $\delta>0$, these branches contain a fold bifurcation, as shown in Fig.~\ref{fig:1d-bif-noise}(b).
The nonexplosive-to-explosive transition from before now occurs in the form of a cusp bifurcation, where the two fold points collide.
We can reach this cusp bifurcation by lowering $\delta$ towards $0$, which will move the fold bifurcations closer towards each other until they meet at a critical value $\delta^\star>0$.
If we lower $\delta$ beyond this point, we enter the situation of Fig.~\ref{fig:1d-bif-noise}(a), where there is only one stable equilibrium and no bifurcations.

However, there is a second way to traverse the cusp bifurcation:
Instead of lowering $\delta$, we can also increase $\varepsilon$.
Intuitively, this results in a \textquote{smoothing} of the curve in Fig.~\ref{fig:1d-bif-noise}(b), until we again end up in a situation like the one in Fig.~\ref{fig:1d-bif-noise}(a).
For this reason, the critical value $\delta^\star$ must depend on $\varepsilon$.

We determine the exact locus of the cusp bifurcation by the following calculation.
Based on the apparent symmetry in Fig.~\ref{fig:1d-bif-noise}, we guess that the cusp will be located at $b=\frac{1}{2}$.
To ease our calculations, we introduce $x\defeq b-\frac{1}{2}$, so that the cusp will be located at $x=0$.
The equation becomes
\[
    \dot x = -2\varepsilon x + \left(\frac{1}{4} - x^2\right) \left(\kappa_1 + \frac{\delta}{2} + \delta x\right) \eqdef f(x; \kappa_1, \delta, \varepsilon) .
\]
For this system to have a cusp bifurcation at $(x^\star; \kappa_1^\star, \delta^\star, \varepsilon^\star)$, we need the conditions
\begin{align*}
    f(x^\star; \kappa_1^\star, \delta^\star, \varepsilon^\star) &= 0\\
    \frac{\pd f}{\pd x} (x^\star; \kappa_1^\star, \delta^\star, \varepsilon^\star) &= 0\\
    \frac{\pd^2f}{\pd x^2} (x^\star; \kappa_1^\star, \delta^\star, \varepsilon^\star) &= 0
\end{align*}
to hold (see e.g.~Sec.~8.2 of \cite{Kuznetsov2023}). The system must also satisfy certain nondegeneracy conditions, but we will deal with those later.

Since we have already guessed $x^\star=0$, we can use the first condition to deduce $\delta^\star = -2\kappa_1^\star$.
The second condition then yields $\kappa_1^\star = -4\varepsilon^\star$.
This already determines the locus of the cusp bifurcation:
\begin{align*}
    x^\star &= 0 \ \left( \text{or equivalently, } b^\star = \frac{1}{2} \right),\\
    \kappa_1^\star &= -4\varepsilon^\star,\\
    \delta^\star &= 8\varepsilon^\star,\\
    \varepsilon^\star &\in (0,\infty).
\end{align*}
It is easy to check that this satisfies the third condition. The nondegeneracy conditions contain a slight subtlety. They are
\begin{align*}
    \frac{\pd^3f}{\pd x^3}(x^\star; \kappa_1^\star, \delta^\star, \varepsilon^\star) &\ne 0 \text{ and }\\
    \left( \frac{\pd f}{\pd p} \frac{\pd^2f}{\pd x\pd q} - \frac{\pd f}{\pd q} \frac{\pd^2f}{\pd x\pd p} \right)(x^\star; \kappa_1^\star, \delta^\star, \varepsilon^\star) &\ne 0,
\end{align*}
where $p$ and $q$ are the two parameters of the cusp bifurcation. Our system, however, has three parameters.
This is not an issue: We can choose any two of $\kappa_1, \delta, \varepsilon$ and the conditions will be satisfied.

In summary, it is also important to point out that our bifurcation analysis has shown that the duplex voter model naturally unfolds the transcritical variant of changing between explosive and non-explosive transitions into a generic codimension-two cusp. This effectively augments and completes the analysis in~\cite{Kuehn2021} for the case of such transitions, where network models without noise were considered.  

\section{Comparison to Simulations}\label{sec:simulations}
In order to test the validity of the mean-field analysis, we simulated our model using the Gillespie algorithm.
Note that all parameter values in this section are given for the model as defined in Sec.~\ref{sec:modeldef}, without the rescaling \eqref{eq:rescaling} for the mean-field.

\begin{figure}
\centering
\subfloat{\begin{overpic}[width=0.45\linewidth]{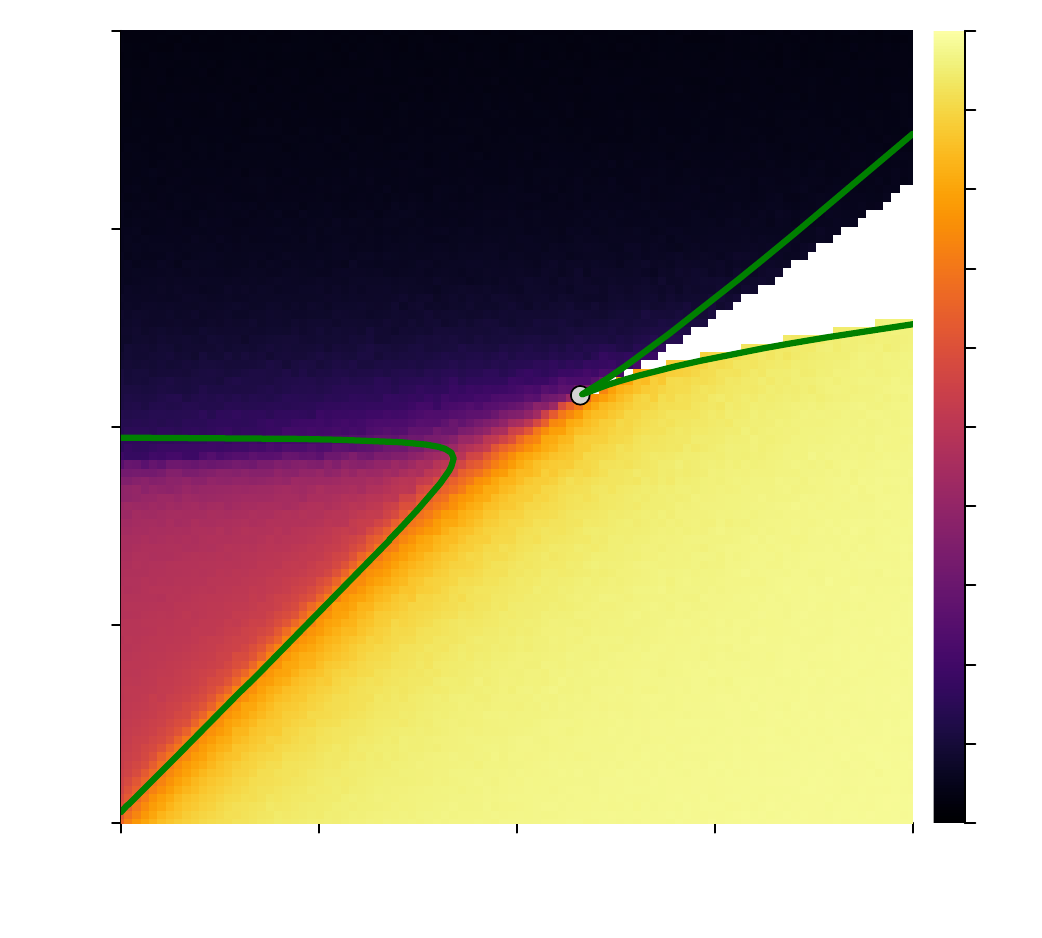}
    \put(50,92){\makebox[0pt]{(a) $\frac{1}{2}(b_1+b_2)$}}
    \put(-4,47.5){$\alpha$}
    \put(2,9.5){\ticksize $0.0$}
    \put(2,28.5){\ticksize $0.5$}
    \put(2,47.5){\ticksize $1.0$}
    \put(2,66.5){\ticksize $1.5$}
    \put(2,85.5){\ticksize $2.0$}

    \put(48,-2){$\delta$}
    \put(5,5){\ticksize $-1.0$}
    \put(23.75,5){\ticksize $-0.5$}
    \put(46.5,5){\ticksize $0.0$}
    \put(65.5,5){\ticksize $0.5$}
    \put(84.5,5){\ticksize $1.0$}
    
    \put(94,10.0){\tiny $0.0$}
    \put(94,17.6){\tiny $0.1$}
    \put(94,25.2){\tiny $0.2$}
    \put(94,32.8){\tiny $0.3$}
    \put(94,40.4){\tiny $0.4$}
    \put(94,48.0){\tiny $0.5$}
    \put(94,55.6){\tiny $0.6$}
    \put(94,63.2){\tiny $0.7$}
    \put(94,70.8){\tiny $0.8$}
    \put(94,78.4){\tiny $0.9$}
    \put(94,86.0){\tiny $1.0$}
    \end{overpic}}
\hfil
\subfloat{\begin{overpic}[width=0.45\linewidth]{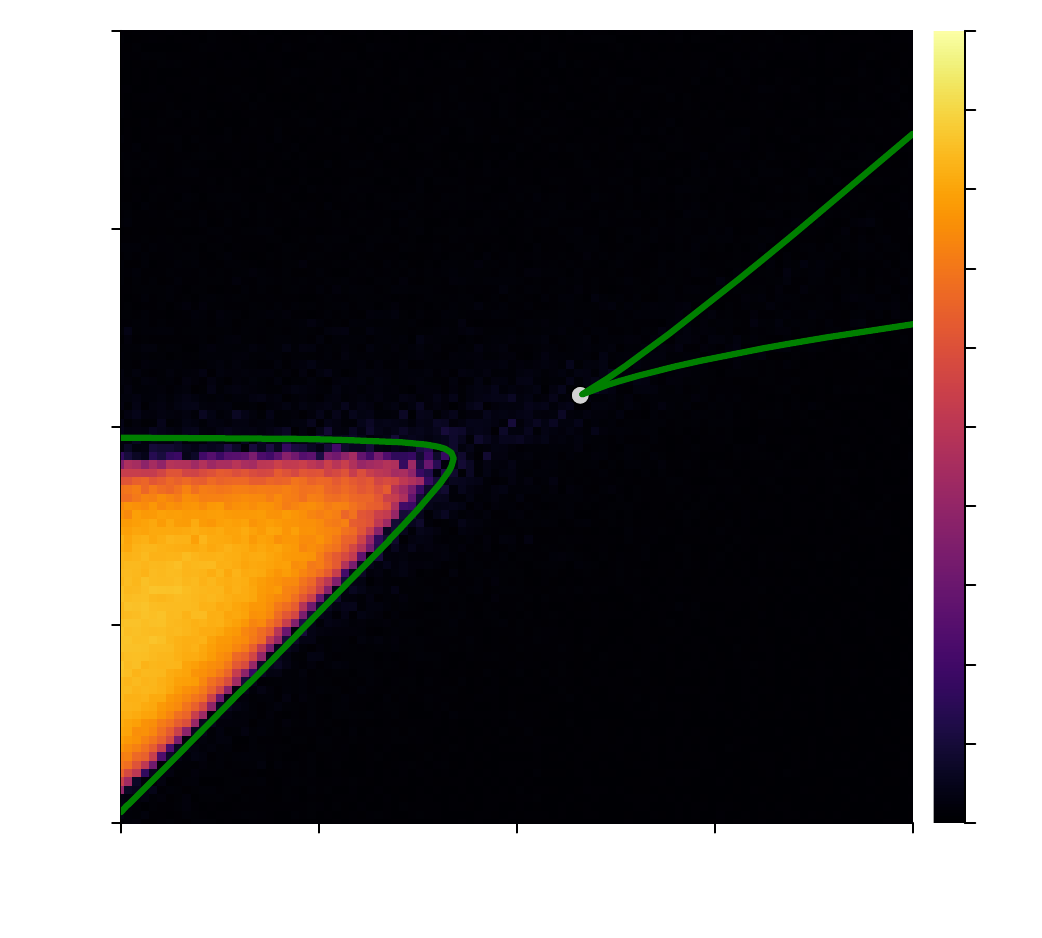}
    \put(50,92){\makebox[0pt]{(b) $|b_1 - b_2|$}}
    \put(-4,47.5){$\alpha$}
    \put(2,9.5){\ticksize $0.0$}
    \put(2,28.5){\ticksize $0.5$}
    \put(2,47.5){\ticksize $1.0$}
    \put(2,66.5){\ticksize $1.5$}
    \put(2,85.5){\ticksize $2.0$}

    \put(48,-2){$\delta$}
    \put(5,5){\ticksize $-1.0$}
    \put(23.75,5){\ticksize $-0.5$}
    \put(46.5,5){\ticksize $0.0$}
    \put(65.5,5){\ticksize $0.5$}
    \put(84.5,5){\ticksize $1.0$}
    
    \put(94,10.0){\tiny $0.0$}
    \put(94,17.6){\tiny $0.1$}
    \put(94,25.2){\tiny $0.2$}
    \put(94,32.8){\tiny $0.3$}
    \put(94,40.4){\tiny $0.4$}
    \put(94,48.0){\tiny $0.5$}
    \put(94,55.6){\tiny $0.6$}
    \put(94,63.2){\tiny $0.7$}
    \put(94,70.8){\tiny $0.8$}
    \put(94,78.4){\tiny $0.9$}
    \put(94,86.0){\tiny $1.0$}
    \end{overpic}}
\caption{
Phase diagram of the mean-field \eqref{eq:2dmf} superimposed on simulation results of the microscopic system for $\varepsilon=0.1$ and $\beta=1$.
The green lines are the bifurcation branches of the mean-field, the gray dot indicates the cusp we calculated in Sec.~\ref{sec:ana-diagonal}.
In (a), we see the projection of the final simulation state onto the diagonal $D$, while (b) shows the distance of the final state to $D$ (scaled by $\sqrt 2$).
The white region in (a) indicates hysteresis.
See text for more information.
}
\label{fig:sim-phase-diagram}
\end{figure}
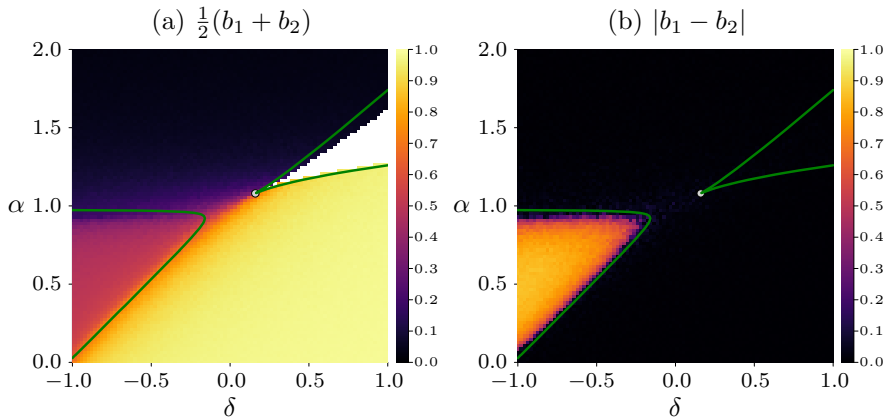

For our first experiment, shown in Fig.~\ref{fig:sim-phase-diagram}, we simulated the model for  a range of values of $(\alpha,\delta)$ and examined the \textquote{final} state at $t=250$.
The other parameters were fixed at $\beta=1,\ \varepsilon=0.1$.
The networks had $5000$ vertices, and the two layers were independent Erdős-Rényi networks with mean degree $k=5.0$.
For each parameter combination $(\alpha,\delta)$, a network was generated and two simulations were performed:
One from a low initial condition $b=(0.01,0.01)$ and one from a high initial condition $b=(0.99,0.99)$.
When the final states of both simulations differed substantially, the corresponding pixel in Fig.~\ref{fig:sim-phase-diagram}(a) was left white, indicating bistability.
Otherwise, it was colored according to the projection of the final state onto the diagonal.
This allows us to see regions corresponding to three (low-B, high-B, bistable) of the four phases we identified in Sec.~\ref{sec:ana-no-noise}.
To also see the symmetry-breaking phase, we show in Fig.~\ref{fig:sim-phase-diagram}(b) the (rescaled) distance of the final simulation state from the diagonal.
The distance would naturally take values in $[0,\frac{1}{\sqrt 2}]$, but we have rescaled it to $[0,1]$ for readability.

For comparison to these simulation results, we have drawn the bifurcation branches of the mean-field \eqref{eq:2dmf}.
These were obtained by numerical continuation using the \textquote{BifurcationKit} Julia package \cite{Veltz2020}.
We also show the cusp point that we have calculated in Sec.~\ref{sec:ana-diagonal}, finding that it agrees with the numerical bifurcation branches.

From the resulting Fig.~\ref{fig:sim-phase-diagram}, we can see that the four phases suggested by the mean-field analysis are indeed found in simulations.
The boundary of the observed symmetry-breaking phase appears consistent with a branch of pitchfork bifurcations in the mean-field system.
This pitchfork bifurcation was invisible to our previous analysis, as it only occurs for $\varepsilon>0$ and is orthogonal to the diagonal.
Regarding the cusp and the corresponding two fold branches in the mean-field, they agree broadly with the region where the simulations exhibit bistability.
However, while the lower fold branch matches the simulations well, the upper fold branch is significantly off.

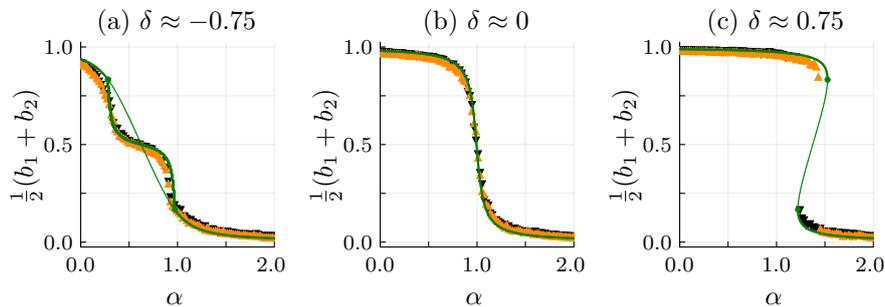
\begin{figure}
\centering
\subfloat{\begin{overpic}[width=0.3\linewidth]{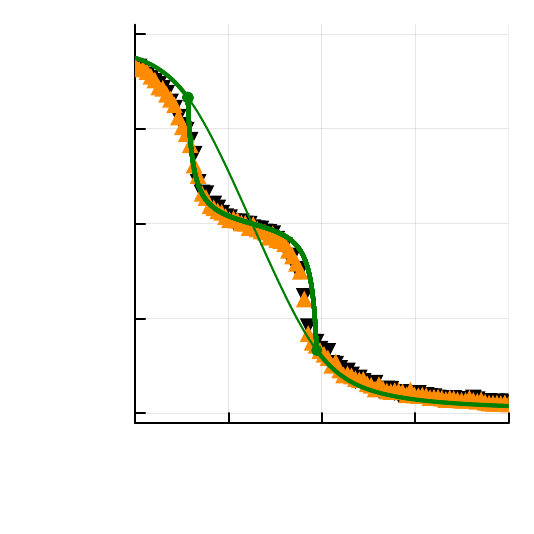}
    \put(60,100){\makebox[0pt]{(a) $\delta\approx-0.75$}}

    \put(0,55){\rotatebox[origin=c]{90}{$\frac{1}{2}(b_1+b_2)$}}
    \put(12,20){\ticksize $0.0$}
    \put(12,55.5){\ticksize $0.5$}
    \put(12,91){\ticksize $1.0$}

    \put(53,0){ $\alpha$}
    \put(20,12){\ticksize $0.0$}
    \put(55,12){\ticksize $1.0$}
    \put(89,12){\ticksize $2.0$}
    \end{overpic}}
\hfil
\subfloat{\begin{overpic}[width=0.3\linewidth]{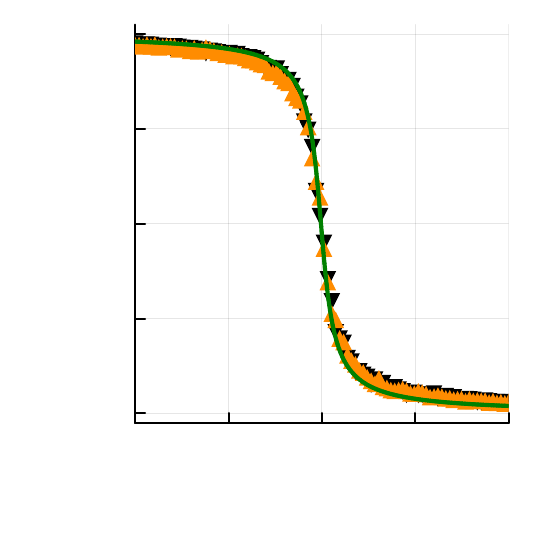}
    \put(60,100){\makebox[0pt]{(b) $\delta\approx0$}}

    \put(0,55){\rotatebox[origin=c]{90}{$\frac{1}{2}(b_1+b_2)$}}
    \put(12,20){\ticksize $0.0$}
    \put(12,55.5){\ticksize $0.5$}
    \put(12,91){\ticksize $1.0$}

    \put(53,0){ $\alpha$}
    \put(20,12){\ticksize $0.0$}
    \put(55,12){\ticksize $1.0$}
    \put(89,12){\ticksize $2.0$}
    \end{overpic}}
\hfil
\subfloat{\begin{overpic}[width=0.3\linewidth]{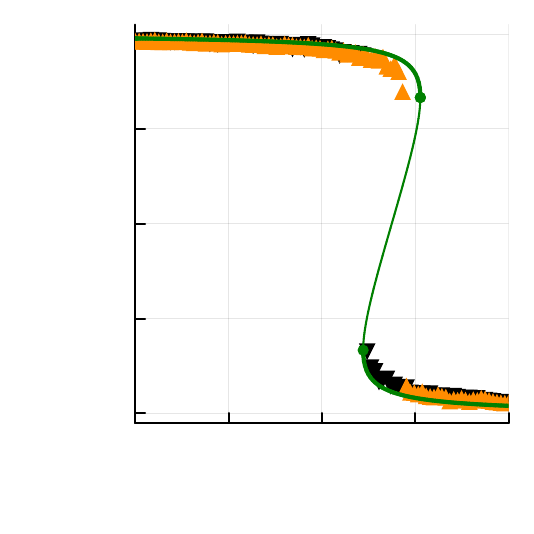}
    \put(60,100){\makebox[0pt]{(c) $\delta\approx0.75$}}

    \put(0,55){\rotatebox[origin=c]{90}{$\frac{1}{2}(b_1+b_2)$}}
    \put(12,20){\ticksize $0.0$}
    \put(12,55.5){\ticksize $0.5$}
    \put(12,91){\ticksize $1.0$}

    \put(53,0){ $\alpha$}
    \put(20,12){\ticksize $0.0$}
    \put(55,12){\ticksize $1.0$}
    \put(89,12){\ticksize $2.0$}
    \end{overpic}}
\caption{
Cross-sections through Fig.~\ref{fig:sim-phase-diagram}(a).
Green lines show the equilibria of the mean-field, where thick lines are stable and thin lines unstable. Green dots mark the bifurcation points.
Triangles are simulation results. The black (orange) triangles are simulations starting from a low (high) initial condition.
}
\label{fig:sim-slices}
\end{figure}

This discrepancy is in fact a symptom of a systematic quantitative difference between the equilibria of the mean-field and those observed in simulations. This can be seen in Fig.~\ref{fig:sim-slices}, where we show three cross-sections through the phase diagram. In fact, it is well-known in related network models that moment-closure schemes can be sensitive to precisely quantify the location of a bifurcation leading to shifts of the bifurcation location~\cite{Demirel2014}. However, one can rigorously prove that most reasonable moment closure schemes actually preserve generic bifurcation types~\cite{KuehnMoelter}. Furthermore, one can employ specialized moment closures near bifurcations to more accurately determine their location~\cite{BoehmeGross}. In summary, our numerical results show quite clearly that the moment closures we selected can capture the mechanisms correctly, and in regions away from bifurcations also extremely accurately.

When interpreting these findings, it's important to keep in mind that the networks used here are ones where we expect our mean-field to perform well:
Erdős-Rényi networks have a degree distribution that is narrowly concentrated around its mean, and they have zero clustering in the large network limit.
Both of these properties make the pair closure \eqref{eq:closure-pair} a reasonable approximation.
In addition, the independence of the two network layers is basically assumed by closure \eqref{eq:closure-other}.
This naturally raises the question of how well the mean-field holds up for other network structures.

For this reason, we have carried out a second experiment, where each network layer was an independent Barabási-Albert network.
These networks feature broad degree distributions, and could therefore cause problems for the closures we used.
The networks had $1000$ vertices, and the mean degree on each layer was $k=8.0$.
We again simulated the model for many values of $(\alpha,\delta)$ until $t=250$.
The other parameters were again fixed at $\beta=1,\ \varepsilon=0.1$.

We have also performed a third experiment, where the two network layers were identical square lattices with periodic boundary.
While this network has a perfectly regular degree distribution with $k=4.0$, it poses other challenges to our closures:
Most importantly, lattices contain many small cycles and the neighborhoods on both layers are the same.
This conflicts with closure \eqref{eq:closure-pair}, which treats all neighbors of a vertex as independent, and with closure \eqref{eq:closure-other}, which additionally treats the states on each layer as independent.
Furthermore, the typical distances in a lattice are large, unlike the \textquote{small worlds} of Erdős-Rényi or Barabási-Albert networks.
In our simulations, we used $32\times 32$ lattices, with $1024$ vertices in total.
We again simulated various values of $(\alpha,\delta)$, for $\beta=1,\ \varepsilon=0.1$.
This time, we ran the simulations until $t=500$ as the model took longer to reach a steady state.

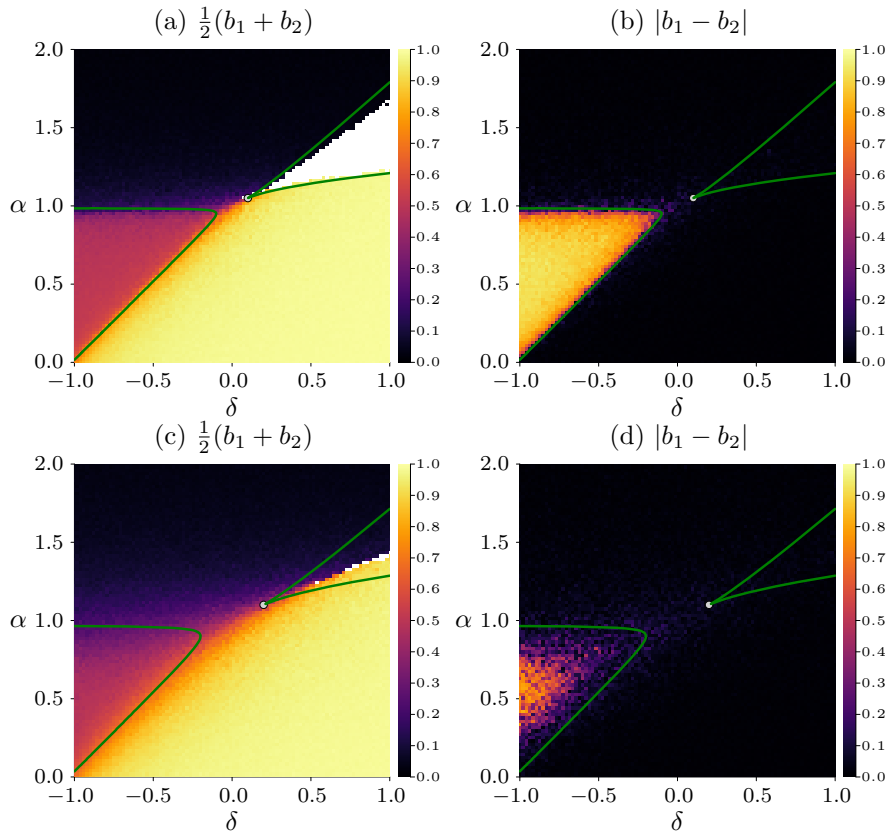
\begin{figure}
\centering
\subfloat{\begin{overpic}[width=0.45\linewidth]{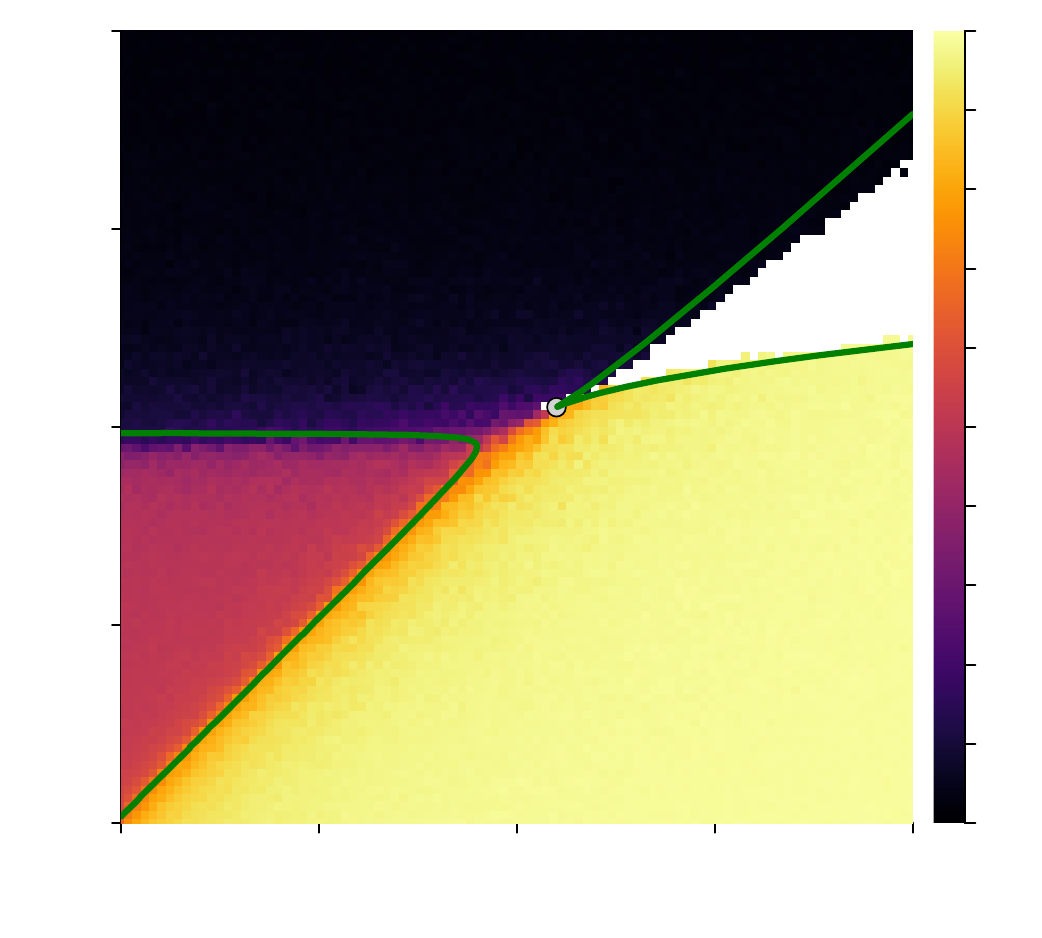}
    \put(50,92){\makebox[0pt]{(a) $\frac{1}{2}(b_1+b_2)$}}
    \put(-4,47.5){$\alpha$}
    \put(2,9.5){\ticksize $0.0$}
    \put(2,28.5){\ticksize $0.5$}
    \put(2,47.5){\ticksize $1.0$}
    \put(2,66.5){\ticksize $1.5$}
    \put(2,85.5){\ticksize $2.0$}

    \put(48,-2){$\delta$}
    \put(5,5){\ticksize $-1.0$}
    \put(23.75,5){\ticksize $-0.5$}
    \put(46.5,5){\ticksize $0.0$}
    \put(65.5,5){\ticksize $0.5$}
    \put(84.5,5){\ticksize $1.0$}
    
    \put(94,10.0){\tiny $0.0$}
    \put(94,17.6){\tiny $0.1$}
    \put(94,25.2){\tiny $0.2$}
    \put(94,32.8){\tiny $0.3$}
    \put(94,40.4){\tiny $0.4$}
    \put(94,48.0){\tiny $0.5$}
    \put(94,55.6){\tiny $0.6$}
    \put(94,63.2){\tiny $0.7$}
    \put(94,70.8){\tiny $0.8$}
    \put(94,78.4){\tiny $0.9$}
    \put(94,86.0){\tiny $1.0$}
    \end{overpic}}
\hfil
\subfloat{\begin{overpic}[width=0.45\linewidth]{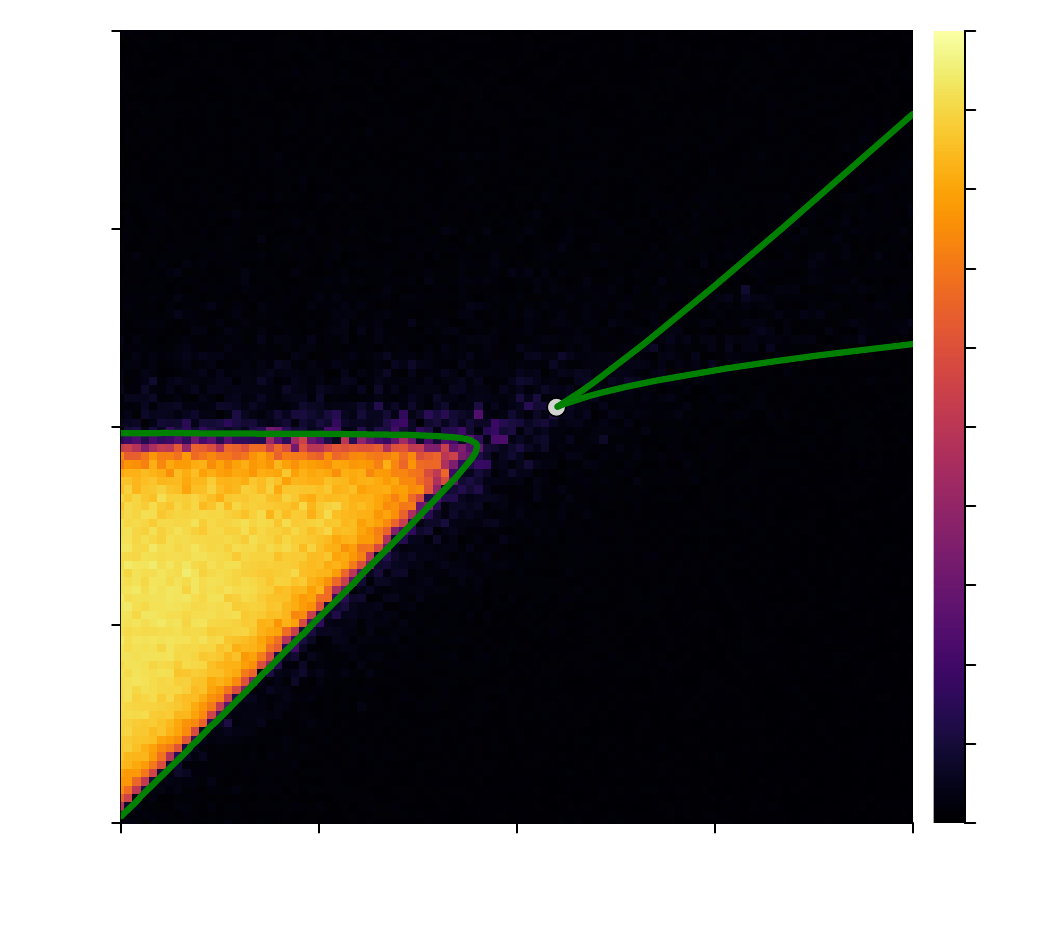}
    \put(50,92){\makebox[0pt]{(b) $|b_1 - b_2|$}}
    \put(-4,47.5){$\alpha$}
    \put(2,9.5){\ticksize $0.0$}
    \put(2,28.5){\ticksize $0.5$}
    \put(2,47.5){\ticksize $1.0$}
    \put(2,66.5){\ticksize $1.5$}
    \put(2,85.5){\ticksize $2.0$}

    \put(48,-2){$\delta$}
    \put(5,5){\ticksize $-1.0$}
    \put(23.75,5){\ticksize $-0.5$}
    \put(46.5,5){\ticksize $0.0$}
    \put(65.5,5){\ticksize $0.5$}
    \put(84.5,5){\ticksize $1.0$}
    
    \put(94,10.0){\tiny $0.0$}
    \put(94,17.6){\tiny $0.1$}
    \put(94,25.2){\tiny $0.2$}
    \put(94,32.8){\tiny $0.3$}
    \put(94,40.4){\tiny $0.4$}
    \put(94,48.0){\tiny $0.5$}
    \put(94,55.6){\tiny $0.6$}
    \put(94,63.2){\tiny $0.7$}
    \put(94,70.8){\tiny $0.8$}
    \put(94,78.4){\tiny $0.9$}
    \put(94,86.0){\tiny $1.0$}
    \end{overpic}}
\vspace{2mm}
\subfloat{\begin{overpic}[width=0.45\linewidth]{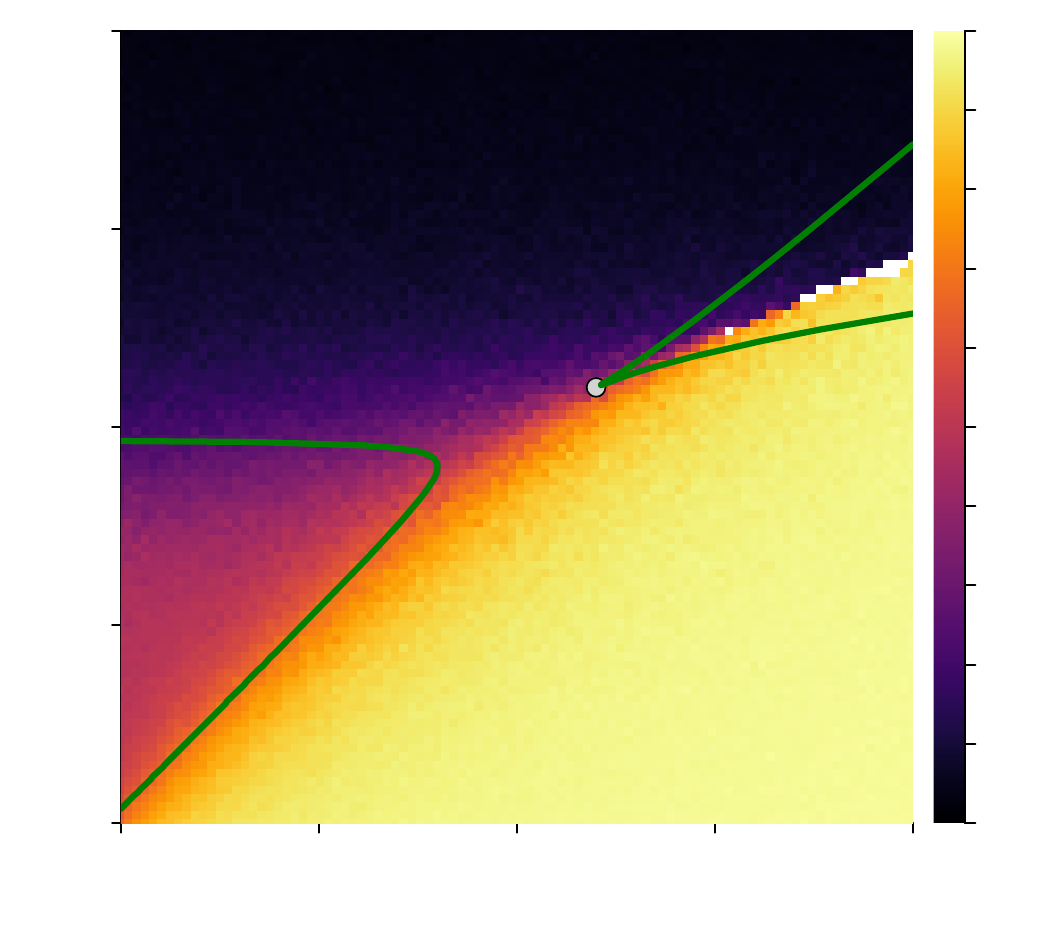}
    \put(50,92){\makebox[0pt]{(c) $\frac{1}{2}(b_1+b_2)$}}
    \put(-4,47.5){$\alpha$}
    \put(2,9.5){\ticksize $0.0$}
    \put(2,28.5){\ticksize $0.5$}
    \put(2,47.5){\ticksize $1.0$}
    \put(2,66.5){\ticksize $1.5$}
    \put(2,85.5){\ticksize $2.0$}

    \put(48,-2){$\delta$}
    \put(5,5){\ticksize $-1.0$}
    \put(23.75,5){\ticksize $-0.5$}
    \put(46.5,5){\ticksize $0.0$}
    \put(65.5,5){\ticksize $0.5$}
    \put(84.5,5){\ticksize $1.0$}
    
    \put(94,10.0){\tiny $0.0$}
    \put(94,17.6){\tiny $0.1$}
    \put(94,25.2){\tiny $0.2$}
    \put(94,32.8){\tiny $0.3$}
    \put(94,40.4){\tiny $0.4$}
    \put(94,48.0){\tiny $0.5$}
    \put(94,55.6){\tiny $0.6$}
    \put(94,63.2){\tiny $0.7$}
    \put(94,70.8){\tiny $0.8$}
    \put(94,78.4){\tiny $0.9$}
    \put(94,86.0){\tiny $1.0$}
    \end{overpic}}
\hfil
\subfloat{\begin{overpic}[width=0.45\linewidth]{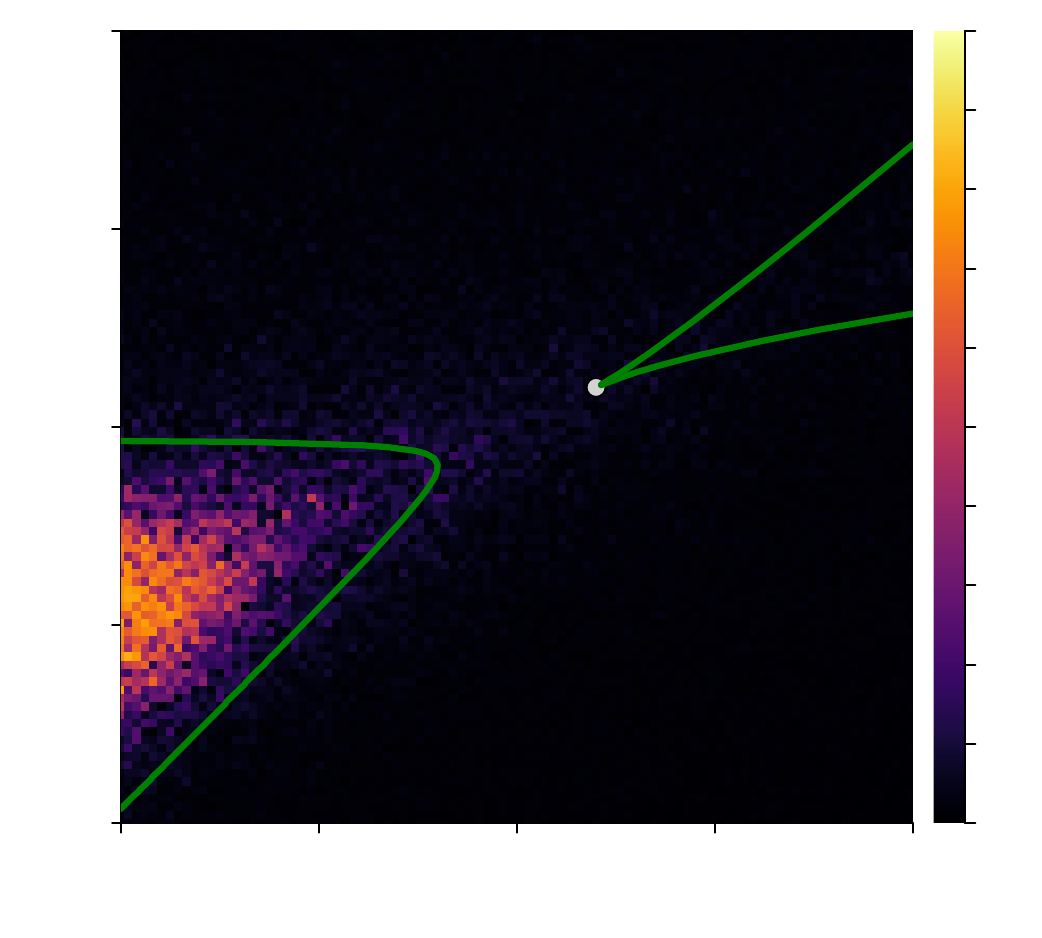}
    \put(50,92){\makebox[0pt]{(d) $|b_1 - b_2|$}}
    \put(-4,47.5){$\alpha$}
    \put(2,9.5){\ticksize $0.0$}
    \put(2,28.5){\ticksize $0.5$}
    \put(2,47.5){\ticksize $1.0$}
    \put(2,66.5){\ticksize $1.5$}
    \put(2,85.5){\ticksize $2.0$}

    \put(48,-2){$\delta$}
    \put(5,5){\ticksize $-1.0$}
    \put(23.75,5){\ticksize $-0.5$}
    \put(46.5,5){\ticksize $0.0$}
    \put(65.5,5){\ticksize $0.5$}
    \put(84.5,5){\ticksize $1.0$}
    
    \put(94,10.0){\tiny $0.0$}
    \put(94,17.6){\tiny $0.1$}
    \put(94,25.2){\tiny $0.2$}
    \put(94,32.8){\tiny $0.3$}
    \put(94,40.4){\tiny $0.4$}
    \put(94,48.0){\tiny $0.5$}
    \put(94,55.6){\tiny $0.6$}
    \put(94,63.2){\tiny $0.7$}
    \put(94,70.8){\tiny $0.8$}
    \put(94,78.4){\tiny $0.9$}
    \put(94,86.0){\tiny $1.0$}
    \end{overpic}}
\caption{
Phase diagram comparison for other networks, with $\varepsilon=0.1,\ \beta=1$.
In (a) and (b), the layers are independent Barabási-Albert networks with mean degree $8.0$.
In (c) and (d), the layers are identical square lattices.
}
\label{fig:sim-phase-diagram-other}
\end{figure}

The results of these two experiments are shown in Fig.~\ref{fig:sim-phase-diagram-other}.
We can see that the mean-field works similarly well on Barabási-Albert networks as it does for Erdős-Rényi.
This could seem surprising, as it is well known that scale-free networks can substantially affect certain types of contagion \cite{PastorSatorras2001}.
But in fact, it is a known property of the biased edge voter model that the degree distribution does not affect its long-time behavior \cite{Antal2006, Sood2008}.
It is apparent that this still holds when one couples two such models.

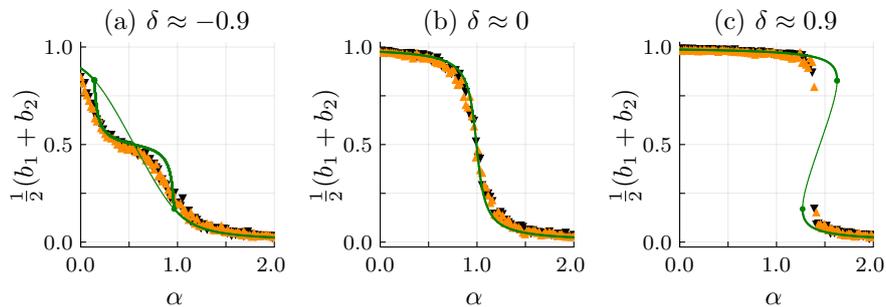
\begin{figure}
\centering
\subfloat{\begin{overpic}[width=0.3\linewidth]{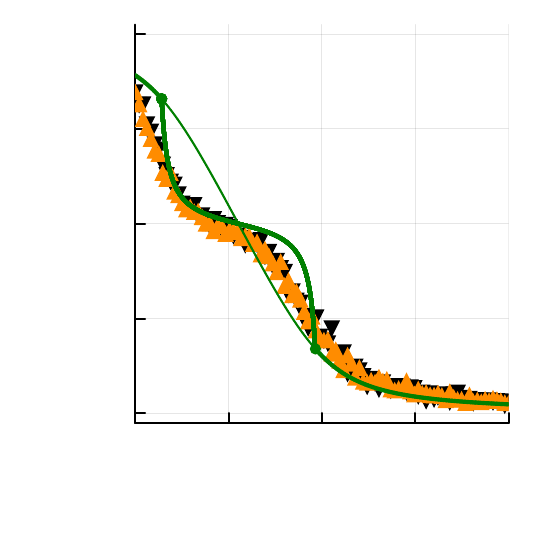}
    \put(60,100){\makebox[0pt]{(a) $\delta\approx-0.9$}}

    \put(0,55){\rotatebox[origin=c]{90}{$\frac{1}{2}(b_1+b_2)$}}
    \put(12,20){\ticksize $0.0$}
    \put(12,55.5){\ticksize $0.5$}
    \put(12,91){\ticksize $1.0$}

    \put(53,0){ $\alpha$}
    \put(20,12){\ticksize $0.0$}
    \put(55,12){\ticksize $1.0$}
    \put(89,12){\ticksize $2.0$}
    \end{overpic}}
\hfil
\subfloat{\begin{overpic}[width=0.3\linewidth]{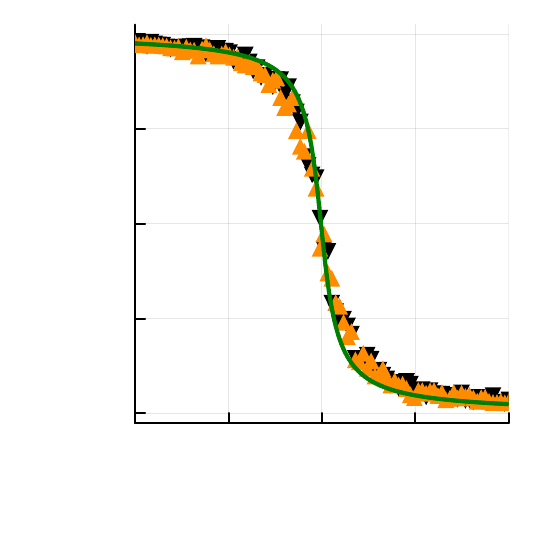}
    \put(60,100){\makebox[0pt]{(b) $\delta\approx0$}}

    \put(0,55){\rotatebox[origin=c]{90}{$\frac{1}{2}(b_1+b_2)$}}
    \put(12,20){\ticksize $0.0$}
    \put(12,55.5){\ticksize $0.5$}
    \put(12,91){\ticksize $1.0$}

    \put(53,0){ $\alpha$}
    \put(20,12){\ticksize $0.0$}
    \put(55,12){\ticksize $1.0$}
    \put(89,12){\ticksize $2.0$}
    \end{overpic}}
\hfil
\subfloat{\begin{overpic}[width=0.3\linewidth]{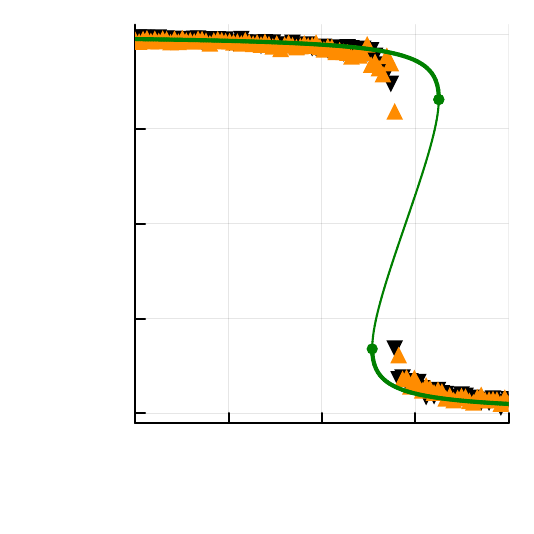}
    \put(60,100){\makebox[0pt]{(c) $\delta\approx0.9$}}

    \put(0,55){\rotatebox[origin=c]{90}{$\frac{1}{2}(b_1+b_2)$}}
    \put(12,20){\ticksize $0.0$}
    \put(12,55.5){\ticksize $0.5$}
    \put(12,91){\ticksize $1.0$}

    \put(53,0){ $\alpha$}
    \put(20,12){\ticksize $0.0$}
    \put(55,12){\ticksize $1.0$}
    \put(89,12){\ticksize $2.0$}
    \end{overpic}}
\caption{
Cross-sections through Fig.~\ref{fig:sim-phase-diagram-other}(c).
Green lines show the equilibria of the mean-field, where thick lines are stable and thin lines unstable. Green dots mark the bifurcation points.
Triangles are simulation results on lattices. The black (orange) triangles are simulations starting from a low (high) initial condition.
}
\label{fig:sim-slices-other}
\end{figure}

However, the mean-field clearly fails to predict the model's behavior on lattices, where the bistable region appears to vanish entirely.
This can be seen very clearly in Fig.~\ref{fig:sim-slices-other}(c), where we show a cross-section through Fig.~\ref{fig:sim-phase-diagram-other}(c).
Additionally, Fig.~\ref{fig:sim-slices-other}(a) indicates that the mean-field also fails to describe the symmetry-breaking phase accurately, though it is not clear if this is merely a substantial quantitative disagreement, or if the bifurcations are qualitatively different.

From these experiments, we conclude that the findings from the mean-field are robust to heterogeneous degree distributions, but are not accurate in all parameter regions in the presence of large distances and local correlations, which can be induced by layer overlap and clustering. However, in such situations it might be better to directly leverage this additional structure, e.g., there are many specialized techniques for contact processes on lattices~\cite{Liggett1999}.

\section{Conclusion}\label{sec:conclusion}
In this paper, we used a mean-field approach to show how catalytic or inhibitory coupling between two individually simple voter models leads to several interesting dynamical phenomena including symmetry breaking and bistability.
In the absence of noise, we were able to characterize all phases of the model and the degenerate bifurcations between them.
We have also seen that the introduction of noise unfolds these degenerate bifurcations into generic ones, and we have pinpointed a resulting cusp bifurcation, from which the region of bistability starts.

Finally, we used simulations of the model on different networks to investigate the extent to which our analysis is valid.
The results show good performance of the mean-field on networks composed of independent Erdős-Rényi or Barabási-Albert layers, in line with the monolayer edge voter model's indifference to broad degree distributions.
By contrast, simulations on identical lattice layers reveal a substantially different behavior than the mean-field predicts.
This is likely caused by correlations between the neighbors of a vertex, induced by many short loops, and between the states of the same vertex on both layers, induced by layer overlap.
Both types of correlations do not satisfy the independence assumptions in the closures that we used to derive the mean-field.
This highlights the need to adapt one's moment closure methods to the network structure at hand, e.g.~by closing only at the triplet level, or by using pair closures that explicitly account for clustering. 

A limitation of this work is the high degree of symmetry in both our model and the considered networks.
By assuming equal average degrees $k_1=k_2$ we focus on a special case, and it stands to reason that the remaining non-generic bifurcation --- the pitchfork that marks the onset of symmetry breaking --- would disappear if one relaxes this assumption.
Additionally, the model is defined around a symmetric $\smat{B\\B}$ coupling, which could be generalized in future work by introducing couplings for other combinations of states.
Or one could study the effects of parasitic and commensal interactions by allowing asymmetric couplings.
Future work could also address similar couplings of other related processes such as the vertex voter model or the invasion process, where the phenomenology will no doubt be even richer.

\section*{Acknowledgments}
This work was supported by the Deutsche Forschungsgemeinschaft (DFG), project number 496237661.

\printbibliography

\end{document}